\documentclass[12pt,preprint,review]{elsarticle}
\usepackage[latin9]{inputenc}
\usepackage{float}
\usepackage{amsmath}
\usepackage{amssymb}
\usepackage{graphicx}

\makeatletter

\providecommand{\tabularnewline}{\\}

\usepackage{amsthm}
\usepackage{lineno}
\usepackage{multirow}

\journal{Pattern Recognition Letters}

\makeatother

\begin{document}
\begin{frontmatter}



\title{Further results on dissimilarity spaces for hyperspectral images RF-CBIR}

\maketitle
\author[label1,label2]{Miguel A. Veganzones\corref{cor1}}
\author[label3]{Mihai Datcu\corref{cor2}}
\author[label1]{Manuel Gra\~{n}a\corref{cor3}}
\address[label1]{Grupo de Inteligencia Computacional, Basque Country University (UPV/EHU), Spain}
\address[label2]{Gipsa-lab, Grenoble-INP, France}
\address[label3]{German Aerospace Center (DLR), Germany}
\cortext[cor1]{E-mail: miguel-angel.veganzones@gipsa-lab.fr - Phone: +33 (0)476 82 6396}
\cortext[cor2]{E-mail: mihai.datcu@dlr.de - Phone: +49 8153 28 1388}
\cortext[cor3]{E-mail: manuel.grana@ehu.es - Phone: +34 943 01 8044}


\begin{abstract}
Content-Based Image Retrieval (CBIR) systems are powerful search tools in image databases that have been little applied to hyperspectral images. Relevance Feedback (RF) is an iterative process that uses machine learning techniques and user's feedback to improve the CBIR systems performance. We pursued to expand previous research in hyperspectral CBIR systems built on dissimilarity functions defined either on spectral and spatial features extracted by spectral unmixing techniques, or on dictionaries extracted by dictionary-based compressors. These dissimilarity functions were not suitable for direct application in common machine learning techniques. We propose to use a RF general approach based on dissimilarity spaces which is more appropriate for the application of machine learning algorithms to the Hyperspectral RF-CBIR. We validate the proposed RF method for hyperspectral CBIR systems over a real hyperspectral dataset.
\end{abstract}

\end{frontmatter}

\linenumbers


\section{Introduction}
\label{sec:intro}

The increasing interest in hyperspectral remote sensing \citep{plaza_recent_2009}
will yield to an exponential growth of hyperspectral data acquisition
in a short time. Most spatial agencies have scheduled the launch of
hyperspectral sensors on satellite payloads such as in EnMAP~\citep{german_spatial_agency_environmental_????}
or PRISMA~\citep{italian_spatial_agency_precursore_????} missions.
That will involve the storage of a huge quantity of hyperspectral
data. The problem of searching through these huge databases using
Content-Based Image Retrieval (CBIR) techniques has not been properly addressed for the case of hyperspectral
images until recently. Recent works on hyperspectral CBIR systems~\citep{grana_endmember-based_????,veganzones_spectral/spatial_2012}
make use of spectral and spectral-spatial dissimilarity functions to compare
hyperspectral images. The spectral and spatial features are extracted by means of spectral unmixing algorithms~\citep{keshava_spectral_2002}. In~\citep{veganzones2012dictionary}, authors define dissimilarity
functions built upon Kolmogorov complexity~\citep{li_introduction_1997}
and its approximation by compression and dictionary distances~\citep{watanabe_new_2002,li_similarity_2004}. Compression-based distances require a high computational cost that make it unaffordable for the definition of CBIR systems. Dictionary distances operate over dictionaries extracted from the hyperspectral images by the off-line application of a lossless dictionary-based compressor such as the Lempel-Ziv-Welch (LZW) compression algorithm~\citep{Welch1984}. In this work we pursued to extend these hyperspectral CBIR systems by using the feedback of the user.

Relevance Feedback (RF) is an iterative process that makes use of the feedback provided by the user to reduce the gap between the low-level feature representation of the images and the high-level semantics of the user's queries~\citep{smeulders_content-based_2000}. Often, the user's feedback comes on the form of a labelling of the previously retrieved images as relevant or irrelevant for the query. The set of labelled images is then used by the CBIR system to adapt the search to the query semantics. If each image is represented by a point in a feature space, the RF with both, positive and negative training examples, becomes a two-class classification problem or an online learning problem in a batch mode \citep{zhou2003}.

Dictionaries and spectral-spatial features extracted from hyperspectral images cannot be directly represented as points in a feature space. Thus, they do not fit easily in feature-based machine learning techniques employed for the definition of RF processes. It is possible to treat dissimilarity functions as kernel functions in order to use them in kernel-based method, for instance in Support Vector Machine (SVM) \citep{shawe-taylor_kernel_2004}. However, these dissimilarity functions do not comply often with valid kernel conditions~\citep{pekalska_kernel_2009}. Authors in~\citep{pekalska_dissimilarity_2005,duin2012} propose the definition
of dissimilarity spaces as an alternative to feature spaces for machine learning. In dissimilarity spaces some data instances are used as reference points named prototypes. The data samples are compared to these prototype instances by some dissimilarity function. Then, for each data sample, the dissimilarities to the prototypes define the data coordinates in a so-called dissimilarity space. Thus, each prototype defines a dimension in this dissimilarity space. The dissimilarity space is analogous to a feature space so, once the data samples are represented as points in the dissimilarity space, all the available potential of machine learning techniques can then be used. 

In this paper we propose the use of dissimilarity spaces to define a RF methodology for hyperspectral CBIR making use of the already available spectral, spectral-spatial and dictionary dissimilarity functions. The use of dissimilarity spaces to define RF processes is scarce on the literature. In \citep{Nguyen2006}, authors propose the use of dissimilarities to prototypes selected by an offline clustering process as the entry to a RF process defined as an one-class classification problem. Authors in \citep{giacinto2003} perform an online prototypes selection instead, where the images retrieved to the user for evaluation are at the same time the prototypes and the training set. The RF process is defined as a new dissimilarity function based on the combination of the database images dissimilarities to the set of prototypes and the prototypes labeling. In \citep{bruno2006}, authors propose different strategies to characterize an image by a feature vector based on the combination of dissimilarities to a set of prototypes. We propose an hyperspectral RF process defined as a two-class classification problem based on dissimilarity spaces. The input to the classifier is a dissimilarity representation defined over the unmixing and dictionary-based hyperspectral dissimilarity functions respect to offline and online selected prototypes.

The paper is divided as follows. In section \ref{sec:dissimilarity} we outline the dissimilarity functions used in the definition of hyperspectral CBIR systems and in section \ref{sec:dis_spaces} we outline the dissimilarity spaces approach. In section \ref{sec:rf_dis_spaces} we introduce the proposed hyperspectral RF process. In section \ref{sec:exper_meth} we define the experimental methodology and in section \ref{sec:results} we comment on the results. Finally, we contribute with some conclusions in section \ref{sec:conclusions}.

\section{Hyperspectral dissimilarity functions}
\label{sec:dissimilarity}

Here, we outline the dissimilarity functions used on the literature to compare hyperspectral images. Firstly, we describe the spectral and spectral-spatial dissimilarity functions defined over the results of a spectral unmixing process. Secondly, we describe the dictionary distance defined over dictionaries extracted from the hyperspectral images by means of lossless dictionary-based compressors. 

\subsection{Unmixing-based dissimilarity functions}
Spectral unmixing pursues the decomposition of an hyperspectral image into the spectral signatures of its main constituents and their corresponding spatial fractional abundances. Most of the unmixing methods are based on the Linear Mixing Model (LMM)~\citep{Keshava2002,Bioucas-Dias2012}. The LMM states that an hyperspectral sample is formed by a linear combination of the spectral signatures of pure materials present in the sample (endmembers), plus some additive noise. Often, the spectral signatures of the materials are unknown, and the set of endmembers must be built by either manually selecting spectral signatures from a spectral library, or by automatically inducing them from the image itself. The latter involves the use of some endmember induction algorithm (EIA). The hyperspectral literature features plenty of such algorithms. Some reviews on the topic can be found in~\citep{plaza_quantitative_2004,veganzones_endmember_2008,Bioucas-Dias2012}. Once the set of endmembers has been induced, their corresponding per-pixel abundances can be estimated by a Least Squares method~\citep{lawson_solving_1974}.

The dissimilarity functions based on the spectral unmixing make use of the spectral and spectral-spatial characterization of the hyperspectral images~\citep{grana_endmember-based_????,veganzones_spectral/spatial_2012}. Given an hyperspectral image, $\mathbf{H}_{\alpha}$, whose pixels are vectors in a $q$-dimensional space, its spectral characterization is defined by the set of endmembers, $\mathbf{E}_{\alpha}=\left\{\mathbf{e}_{1}^{\alpha},\mathbf{e}_{2}^{\alpha}\ldots\mathbf{e}_{m_{\alpha}}^{\alpha}\right\}$, where $m_{\alpha}$ denotes the number of induced endmembers from the $\alpha$-th image. The spectral-spatial characterization is defined as the tuple $\left(\mathbf{E}_{\alpha},\mathbf{\Phi}_{\alpha}\right)$, where $\mathbf{\Phi}_{\alpha}=\left\{ \phi_{1}^{\alpha},\phi_{2}^{\alpha},\ldots,\phi_{m_{\alpha}}^{\alpha}\right\}$ is the set of fractional abundance maps resulting from the unmixing process. To implement this approach, an EIA is first used to induce the endmembers from the image and then, their respective fractional abundances are estimated by a Least Squares Unmixing algorithm. 

In order to compute the unmixing-based dissimilarities, the Spectral Distance Matrix (SDM), $D_{\alpha,\beta}$, between two given hyperspectral images, $H_{\alpha}$ and $H_{\beta}$, has first to be computed. The SDM is the matrix $D_{\alpha,\beta}=\left[d_{ij}\right]$, $i=1,\ldots,m_{\alpha}$, $j=1,\ldots,m_{\beta}$, whose elements $d_{ij}$ are the pairwise distances between the endmembers $\mathbf{e}_{i}^{\alpha},\mathbf{e}_{j}^{\beta}\in\mathbb{R}^{q}$ of each image. The spectral distance function $d:\mathbb{R}^{q}\times\mathbb{R}^{q}\rightarrow\mathbb{R}^{+}$ is often the angular pseudo-distance:
\begin{equation}
d\left({\mathbf{e}}_{i},{\mathbf{e}}_{j}\right) = \cos^{-1}\frac{\mathbf{e}_{i}\mathbf{e}_{j}}{\left\Vert{\mathbf{e}}_{i}\right\Vert \left\Vert{\mathbf{e}}_{j}\right\Vert}\label{eq:angledistance}.
\end{equation}
The Spectral dissimilarity~\citep{grana_endmember-based_????} is then given by:
\begin{equation}\label{eq:spectral_dissimilarity}
s_{\mathbf{E}}\left(\mathbf{H}_{\alpha},\mathbf{H}_{\beta}\right) = \left\Vert \mathbf{m}_{r}\right\Vert +\left\Vert \mathbf{m}_{c}\right\Vert ,
\end{equation} 
where $\left\Vert\mathbf{m}_{r}\right\Vert$ and $\left\Vert\mathbf{m}_{c}\right\Vert$ are the Euclidean norms of the vectors of row and column minimal values of the SMD, respectively. The Spectral-Spatial dissimilarity~\citep{veganzones_spectral/spatial_2012} is given by:
\begin{equation}\label{eq:spectralspatial_dissimilarity}
s_{\mathbf{E},\mathbf{\Phi}}\left(\mathbf{H}_{\alpha},\mathbf{H}_{\beta}\right)= \sum_{i=1}^{m_\alpha}\sum_{j=1}^{m_\beta}r_{ij}d_{ij},
\end{equation}
where $d_{ij}$ is the aforementioned spectral distance and $r_{ij}$ is the significance associated to $d_{ij}$. The significance matrix $R_{\alpha,\beta}=\left[r_{ij}\right]$, $i=1,\ldots,m_{\alpha}$, $j=1,\ldots,m_{\beta}$ is calculated on base to the normalized average abundances $\bar{\mathbf{\Phi}}_{\alpha}$ and $\bar{\mathbf{\Phi}}_{\beta}$ by the most similar highest priority (MSHP) principle~\citep{li_irm_2000}.
 
\subsection{Dictionary-based dissimilarity functions}
Given a signal $x$, a dictionary-based
compression algorithm looks for patterns in the input sequence from
signal $x$. These patterns, called \emph{words}, are subsequences
of the incoming sequence. The compression algorithm result is a set
of unique words called \emph{dictionary.} The dictionary extracted
from a signal $x$ is hereafter denoted as $D\left(x\right)$, with
$D\left(\lambda\right)=\emptyset$ only if $\lambda$ is the empty
signal. The Normalized Dictionary Distance (NDD)~\citep{macedonas_dictionary_2008} is given by:
\begin{equation}
s_{\textrm{NDD}}\left(x,y\right)=\frac{D\left(x\cup y\right)-\min\left\{ D\left(x\right),D\left(y\right)\right\} }{\max\left\{ D\left(x\right),D\left(y\right)\right\} },\label{eq:NDD}
\end{equation}
where $D\left(x\cup y\right)$ and $D\left(x\cap y\right)$ respectively
denote the union and intersection of the dictionaries extracted from
signals $x$ and $y$. The NDD is a normalized admissible
distance satisfying the metric inequalities. Thus, it results in
a non-negative number in the interval $\left[0,1\right]$, being zero
when the compared signals are equal and increasing up to one as the
signals are more dissimilar.

\section{Dissimilarity spaces}
\label{sec:dis_spaces}

The dissimilarity space is a vector space in which the dimensions are defined by dissimilarity vectors measuring pairwise dissimilarities between individual objects and reference objects (prototypes) \citep{duin2012}. Given a set of prototypes $ \textbf{P}=\left\lbrace p_1,\ldots,p_{r}\right\rbrace$, where $ r $ denotes the number of prototype objects on $ \textbf{P} $, and a set of objects $ \textbf{X}=\left\lbrace x_1,\ldots,x_{N}\right\rbrace $, , where $ N $ denotes the number of individual objects on $ \textbf{X} $, the dissimilarity representation $ D\left(\textbf{X},\textbf{P}\right) $ is a data-dependent mapping $ D\left(\cdot,\textbf{P}\right):\textbf{X}\rightarrow\Re^r $ from a set of objects $ \textbf{X} $ to the dissimilarity space specified by the prototypes set $ \textbf{P} $. Each dimension in the dissimilarity space corresponds to a dissimilarity to a prototype object, $ D\left(\textbf{X},p_i\right) $. The dissimilarity representation $ D\left(\textbf{X},\textbf{P}\right) $ is thus defined as a $ N \times r $ dissimilarity matrix, where each object $ \textbf{x}\in\textbf{X} $ is described by a vector of dissimilarities $ \textbf{s}_x=D\left(x,\textbf{P}\right) = \left[s\left(x,p_1\right),\ldots,s\left(x,p_r\right)\right] $. The pairwise dissimilarity function $ s\left(x,p_i\right) $ is not required to be metric and can be defined \textit{ad-hoc} for the given prototype. The dissimilarity space is a vector space equipped with an inner product and an Euclidean metric. Thus, the vector of dissimilarities to the set of prototypes, $ \textbf{s}_x $, can be interpreted as a feature, allowing the use of machine learning techniques commonly defined over feature spaces.

\section{Relevance feedback by dissimilarity spaces}
\label{sec:rf_dis_spaces}

\begin{figure}
\begin{centering}
\includegraphics[width=1\columnwidth]{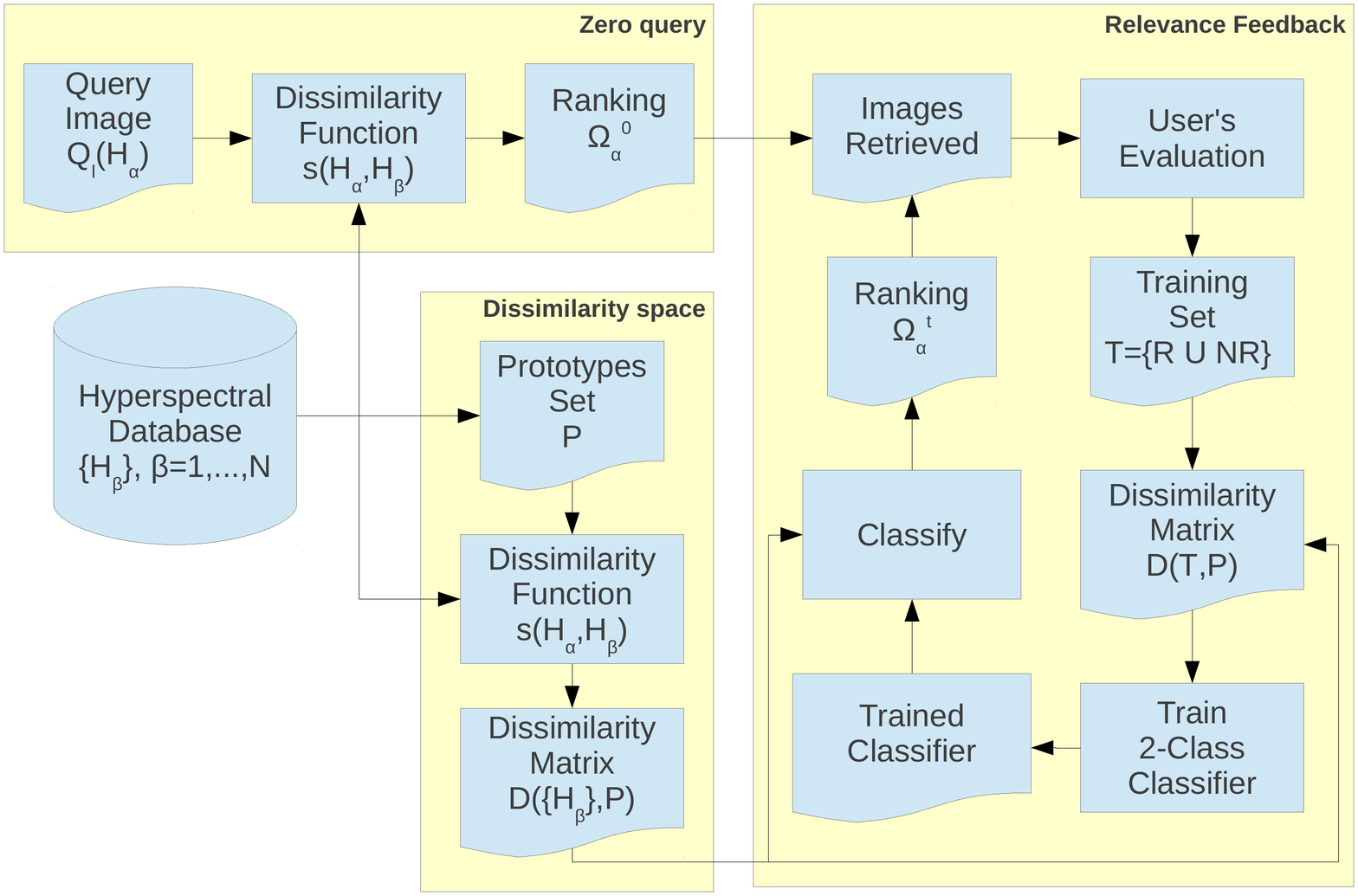} 
\par\end{centering}
\caption{\label{fig:CBIR-system-diagram}CBIR system diagram with the proposed
relevance feedback by dissimilarity spaces approach.}
\end{figure}

The use of dissimilarity spaces allows one to use the previously mentioned hyperspectral dissimilarity functions to define a RF process based on conventional machine learning techniques. The proposed hyperspectral RF process follows the general approach in \citep{giacinto2003,bruno2006,Nguyen2006} and it is depicted in Fig.\ref{fig:CBIR-system-diagram}. First, the user defines a \textit{zero-query} by feeding the system with some positive sample. Next, an initial ranking is obtained comparing the database images to the query sample by some hyperspectral dissimilarity function and some images are retrieved for user's evaluation. Then, the user labels the images retrieved by the system, a set of prototype images is selected and the RF process starts.  We follow by describing the zero-query and the relevance feedback processes in detail, and then we discuss on the prototypes selection and the selection of the images retrieved by the system for evaluation.

\subsection{Zero query}

First, a query $Q_{l}\left(H_{\alpha}\right)$ is defined following
the query-by-image approach. $H_{\alpha}$ denotes the hyperspectral image selected as the query and $l\in\mathbb{Z}^{+}$, named the scope of the query, denotes the number of images that should be retrieved by the system. Every image $H_{\beta}$ in the dataset is compared to the query image by some hyperspectral dissimilarity function, $s\left(H_{\alpha},H_{\beta}\right)$. The dissimilarities to the query image are represented as a vector $\mathbf{s}_{\alpha}=\left[s_{\alpha,1},\ldots,s_{\alpha,N}\right]$, where $N$ is the number of images in the dataset and $s_{\alpha,\beta}$ is the dissimilarity between the query image $H_{\alpha}$ and the dataset image $H_{\beta}$, with $\beta=1,\ldots,N$. Then, we sort the components of $\mathbf{s}_{\alpha}$ in increasing order, and the resulting shuffled image indexes constitute the zero ranking $\Omega_{\alpha}^{0}=\left[\omega_{q}\in\left\{ 1,\ldots,N\right\} \right]$, $q=1,\ldots,N$, so that $s_{\alpha,\omega_{q}}\leq s_{\alpha,\omega_{q+1}}$. Then, some selection criterion is followed to select $l$ images from the zero ranking and retrieve them for user's evaluation. The user labels these images as relevant or non-relevant for the query. The set of relevant images, denoted as $ R $, and the set of non-relevant images, denoted as $ NR $, form the training set, $ T=\left\lbrace R \cup NR \right\rbrace $, with which the relevance feedback process starts.

\subsection{Relevance feedback}

We propose a RF process defined as a two-class problem where the classes are the set of relevant (positive class) and the set of irrelevant (negative class) images respect to the query. The input to the two-class classifier is a feature vector composed of the dissimilarity values computed from a given image respect to each of the images of the prototypes set. The output of the classifier should be an scalar representing any measure of an image identification with the positive class respect to the negative class, for instance a class probability. The classifier outputs are ordered to define a ranking of the database images respect to the user's query. Finally, the ranking is used to select some database images that will be retrieved for the user's evaluation and so, proceed with a new RF iteration. Thus, the RF process is divided in two steps, a training phase and a testing phase.

\subsubsection{Training phase}

Let $\textbf{P}=\left\lbrace H_{p_i}\right\rbrace_{i=1}^{r}$ be the set of prototypes where $ p_i $ is an index pointing to a database image and $ r $ is the number of prototype instances. Let $ \textbf{T}=\left\{ H_{q_j}\right\} _{j=1}^{t} $ be the set of training samples where $ q_j $ is an index pointing to a database image, $ t $ denotes the number of training samples and each image $H_{q_j}$ has been labelled as belonging to the positive class, $\mathcal{C}^{+}$, or to the negative class, $\mathcal{C}^{-}$. Then, the system calculates the $ t \times r $ dissimilarity matrix $D\left(\textbf{T},\textbf{P}\right)=\left[s\left( H_{q_j},H_{p_i}\right)\right]$, $ j=1,\ldots,t $, $ i=1,\ldots,r $; using some given hyperspectral dissimilarity function $ s\left(\cdot,\cdot\right) $. The rows of $D\left(\textbf{T},\textbf{P}\right)$ are the geometrical coordinates of the training samples in the dissimilarity space defined by the set of prototypes, and
would be used as feature vectors to train the two-class classifier.

\subsubsection{Testing phase}

For each image $H_{\beta}$ in the dataset we calculate the dissimilarity vector $\textbf{s}_{\beta}=D\left(H_\beta,\textbf{P}\right)=\left[s\left(H_\beta,H_{p_i}\right)\right]_{i=1}^{r}$, given the hyperspectral dissimilarity function $ s\left(\cdot,\cdot\right) $. The dissimilarity vector, $\mathbf{s}_{\beta}$,
represents a point in the dissimilarity space and is used as the input to the trained classifier. The classifier will return an scalar, $ c_\beta $, measuring the probability or the degree of inclusion of the image $H_{\beta}$ respect to the query class $\mathcal{C}^{+}$. An image $ H_k $ having a classification value higher than an image $ H_l $, that is  $ c_k \geq c_l $, should be ranked in a better position.
The values obtained by the classifier for all the images in the dataset are then represented as a vector $\mathbf{c}_{\alpha}=\left[c_1,\ldots,c_N\right]$,
where $N$ is the number of images in the dataset. The vector of classification values $ \mathbf{c}_{\alpha} $ is sorted in decreasing order and the resulting shuffled image indexes constitute the ranking $\Omega_{\alpha}^{t}=\left[\omega_{q}^{t}\in\left\{ 1,\ldots,N\right\} \right]$, $q=1,\ldots,N$, so that $c_{\omega_{q}^{t}}\geq c_{\omega_{q+1}^{t}}$. The superscript $t$ in $\Omega_{\alpha}^{t}$ denotes the iteration in turn on the RF process, being $t$ a positive integer, $t>0$. The ranking serves to select some images that are retrieved to the user for evaluation, and then included in the training set. The RF process ends when the user is satisfied, a maximum number of iterations, $t_{\textrm{max}}$, is achieved, or no new images are being incorporated to the training set.

\subsection{Prototypes selection}
The general RF process depicted in Fig.\ref{fig:CBIR-system-diagram} requires of a set of prototypes. We distinguish between two criteria to build the prototypes set, an offline selection and an online selection. In the former, the prototypes are \textit{a priori} representative subset of the images in the database. A common procedure is to perform a clustering and keep the centres of the clusters as the prototypes. This criterion could lead to a dramatical reduction in the computational costs of the CBIR system, but on the other hand it defines a fixed set of prototypes for all the possible queries, limiting the adaptability of the CBIR system. The later builds the set of prototypes during the RF process. In each iteration some images are retrieved to the user for evaluation and then included on the training set. These same images or a subset of them are also used as prototypes. This allows to adapt the set of prototypes to the query. However, it increases the computational burden.

\subsection{Image retrieval}
A key aspect of RF-CBIR systems is the criterion to select from a given ranking those images that will be retrieved to the user for evaluation. Let $ l $ denote the scope of the query, that is, the number of images that should be retrieved to the user. If the criterion is to return the best $ l $ images given by the $ l $ best ranked images on the database, is likely that the training set is biased towards the positive class. So, a better criterion seems to retrieve the $ l/2 $ best images and the $ l/2 $ worst images, hereafter denoted as the \textit{Best-Worst} (BW) criterion. However, the best and worst images are not necessarily the most informative ones. The active learning paradigm \citep{libsvm} states that the most ambiguous images, those that are close to the class boundaries, are the most informative. Thus, the \textit{Active Learning} (AL) criterion will return the $ l/2 $ most ambiguous images labelled as belonging to the positive class, and the $ l/2 $ most ambiguous images labelled as belonging to the negative class.

\section{Experimental methodology}
\label{sec:exper_meth}

\subsection{Dataset}

The hyperspectral HyMAP data was made available from HyVista Corp.
and German Aerospace Center's (DLR) optical Airborne Remote Sensing
and Calibration Facility service%
\footnote{http://www.OpAiRS.aero%
}. The scene corresponds to a flight line over the facilities of the
DLR center in Oberpfaffenhofen (Germany) and its surroundings, mostly
fields, forests and small towns. The data cube has $2878$ lines,
$512$ samples and $125$ spectral bands. We have removed non-informative
bands due to atmospheric absorption and $113$ spectral bands remained.

\begin{figure}
\begin{centering}
\begin{tabular}{ccccc}
\includegraphics[width=0.15\columnwidth]{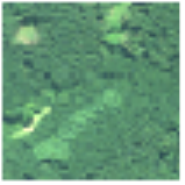}  & \includegraphics[width=0.15\columnwidth]{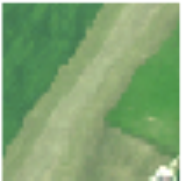}  & \includegraphics[width=0.15\columnwidth]{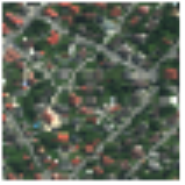}  & \includegraphics[width=0.15\columnwidth]{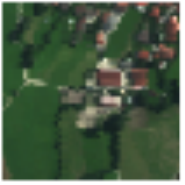}  & \includegraphics[width=0.15\columnwidth]{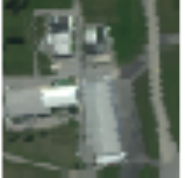}\tabularnewline
(a)  & (b)  & (c)  & (d)  & (e)\tabularnewline
\end{tabular}5 most ambiguous positive and negative instances
\par\end{centering}
\caption{\label{fig:Examples-patches}Examples of the five categories patches:
(a) Forests, (b) Fields, (c) Urban Areas, (d) Mixed, (e) Others.}
\end{figure}

We cut the scene in patches of $64\times64$ pixels size for a total
of 360 patches forming the hyperspectral database used in the experiments.
We grouped the patches by visual inspection in five rough categories.
The three main categories are 'Forests', 'Fields' and 'Urban Areas',
representing patches that mostly belong to one of this categories.
A 'Mixed' category was defined for those patches that presented more
than one of the three main categories, being not any of them dominant.
Finally, we defined a fifth category, 'Others', for those patches
that didn't represent any of the above or that were not easily categorized
by visual inspection. The number of patches per category are: (1)
Forests: 39, (2) Fields: 160, (3) Urban Areas: 24, (4) Mixed: 102,
and (5) Others: 35. Figure \ref{fig:Examples-patches} shows examples
of the five categories patches.

\subsection{Methodology}

We test the use of the proposed hyperspectral RF-CBIR using the unmixing and dictionary-based hyperspectral dissimilarity functions. For the unmixing-based dissimilarities, the spectral \eqref{eq:spectral_dissimilarity} and the spectral-spatial \eqref{eq:spectralspatial_dissimilarity} dissimilarity functions, we conduct for each image in the database an unmixing process in order to obtain the set of induced endmembers and their corresponding fractional abundances. In order to do that we use the Vertex Component Analysis (VCA) \citep{nascimiento2005} endmember induction algorithm and a partially constrained least squares unmixing (PCLSU) \citep{lawson_solving_1974} algorithm. As VCA is an stochastic algorithm we perform 20 independent runs for each image and we keep the one with the lowest averaged root squared mean reconstruction error:
\begin{equation}
\epsilon\left(H,\hat{H}\right) = \frac{1}{M} \sum_{i=1}^{M} \sqrt {\frac{1}{q} \sum_{j=1}^{q} {\left( H_{i}^{\left(j\right)} - \hat{H}_{i}^{\left(j\right)} \right)}^2}
\end{equation}
where $ H_{i}^{\left(j\right)} $ denotes the $ j $-th band value of the $ i $-th pixel in the hyperspectral image $ H $ and $ \hat{H}=\mathbf{\Phi}\textbf{E} $ is the hyperspectral image reconstructed by the set of induced endmembers $ \textbf{E} $ and their corresponding fractional abundances $ \mathbf{\Phi} $.
For the Normalized Dictionary Distance \eqref{eq:NDD}, we first convert each hyperspectral image to a text string in two ways, using the average of the spectral bands and band-by-band. For the former, we calculate the mean of each hyperspectral pixel along the spectral bands. For the later we transform each spectral band independently. In both cases we traverse the image in a zig-zag way. The averaged band transformation incurs in a big lost of spectral information compared to the band by band transformation, but by contrast it yields to a more compact dictionary and so, to speed up the NDD computation.

Thus, we compare the use of the four hyperspectral dissimilarities, the Spectral, the Spectral-Spatial, the Averaged Band NDD and the Band-by-Band NDD, in the RF process respect to their use in the zero-query. In order to do that, we run independent retrieval experiments over the HyMAP dataset. Each of the $360$ patches was \textit{a priori} labelled as belonging to one of the five categories defined above. The query is a categorical search, where the images belonging to the same category than the query image form the positive class and the remaining ones form the negative class. We perform an independent search for each of the $360$ patches. Thus, user's evaluation was not required and the experiment was fully automatized. The maximum number of iterations on the retrieval feedback process was set to $t_{\textrm{max}}=5$.

For the RF process we compare the use of a $ k $-NN classifier and a two-class SVM classifier with a radial basis kernel. The $ k $-NN classifier does not require of a training phase and returns the fraction of the $k$ most similar images in the training set respect to the query image belonging to the positive class, that is, $c = \frac{\sum_{i=1}^{k} I\left(H_i,H_\alpha\right)}{k}$, where $ I $ denotes an indicator function returning 1 if the two images belong to the same class, and 0 otherwise. The SVM classifier outputs the probability that the tested image belongs to the positive class. The parameters of the SVM classifier where selected using a 5-fold cross validation. For the $k$-NN the \textit{knnclassify} MATLAB function was used. For the SVM, we used the $C$-SVM classifier of the LIBSVM \citep{libsvm} library.

We also compare the use of online and offline prototypes selection processes. For the offline prototypes selection process we performed a hierarchical segmentation using each of the four hyperspectral dissimilarity functions and we keep 10 clusters. Then, for each cluster $ \zeta $ we selected the image $ H_{\zeta}^o $ minimizing the averaged distance to the rest of images grouped into the same cluster:
\begin{equation}
H_{\zeta}^o = \arg\min_{i} \frac{1}{\left|\zeta\right|}\sum_{H_{j}\in\zeta} s\left(H_i,H_j\right)
\end{equation}
where $\left|\zeta\right|$ denotes the cardinality of the cluster $\zeta$.

Finally, we compare the results obtained using three different criteria to select the images to be retrieved to the user for evaluation: the BW criterion, the AL criterion and a combination of both, BW+AL. For the BW criterion the system retrieves the 5 best and worst ranked images in the database. For the AL criterion the system retrieves the 5 most ambiguous positive and negative instances, that is, the ones closed to the class boundary on each side. For both, BW and AL criteria, the scope is then $l=10$. For the BW+AL criterion the system returns the 3 best and worst ranked images, and the 3 most ambiguous positive and negative instances, for a total scope of $l=12$.

\subsection{Performance measures}

Evaluation metrics from information retrieval field have been adopted to evaluate CBIR systems quality. The two most used evaluation measures are \emph{precision} and \emph{recall} \citep{smeulders_content-based_2000,daschiel_information_2005}. Precision, $p$, is the fraction of the returned images that are relevant to the query. Recall, $q$, is the fraction of retrieved relevant images respect to the total number of relevant images in the database according to \textit{a priori} knowledge. If we denote $L$ the set of returned images and $R$ the set of all the images relevant to the query, then $p=\frac{\left|L\cap R\right|}{\left|L\right|}$ and $r=\frac{\left|L\cap R\right|}{\left|R\right|}$. Precision and recall follow inverse trends when considered as functions of the scope of the query. Precision falls while recall increases as the scope increases. Thus, precision and recall measures are usually given as precision-recall curves for a fixed scope. To evaluate the overall performance of a CBIR system, the Average Precision and Average Recall are calculated over all the query images in the database. For a query of scope $l$, these are defined as: 
\begin{equation}
\bar{p}_{l}=\frac{1}{N}\sum_{\alpha=1}^{N}p_{l}(H_{\alpha})\label{eq:averaged_precision}
\end{equation}
 and
\begin{equation}
\bar{r}_{l}=\frac{1}{N}\sum_{\alpha=1}^{N}r_{l}(H_{\alpha}).\label{eq:averaged recall}
\end{equation}

The Normalized Rank \citep{muller_performance_2001} was used to summarize the system performance into an scalar value. The normalized rank for a given image query, denoted as $\textrm{Rank}\left(H_{\alpha}\right)$, is defined as: 
\begin{equation}
\textrm{Rank}\left(H_{\alpha}\right)=\frac{1}{NN_{\alpha}}\left(\sum_{i=1}^{N_{\alpha}}\Omega_{\alpha}^{i}-\frac{N_{\alpha}\left(N_{\alpha}-1\right)}{2}\right),\label{eq:Rank}
\end{equation}
where $N$ is the number of images in the dataset, $N_{\alpha}$ is the number of relevant images for the query $H_{\alpha}$, and $\Omega_{\alpha}^{i}$ is the rank at which the $i$-th image is retrieved. This measure is $0$ for perfect performance, and approaches $1$ as performance worsens, being $0.5$ equivalent to a random retrieval. We calculated the $\textrm{Rank}\left(H_{\alpha}\right)$ for each of the images in the dataset and then we calculated the average normalized rank (ANR):
\begin{equation}
ANR=\frac{1}{N}\sum_{\alpha=1}^{N}Rank\left(H_{\alpha}\right).\label{eq:ANR}
\end{equation}

\section{Results}
\label{sec:results}

Tables \ref{tab:ANR_Fields}-\ref{tab:ANR_Urbanareas} show the ANR \eqref{eq:ANR} values of the comparing hyperspectral dissimilarities, using the proposed RF-CBIR respect to the zero-query, for the Forest, Fields and Urban areas categorical queries respectively. We run the experiments using different values of $k$ for the $k$-NN classifier, but we only show the results using $k=7$ as in general it outperforms the other $k$ values. The ANR results correspond to the ranking obtained in the fifth RF iteration. In general, the hyperspectral RF process yields to better ANR results than the zero query for the four compared hyperspectral dissimilarity functions. The online prototype selection leads to better results than the offline selection, and so it does the $7$-NN classifier compared to the SVM classifier. The use of AL for the image retrieval selection outperforms the BW criterion, and often the combination of both, BW+AL. As it was expected, the results using the Band-by-Band NDD and the Spectral-Spatial dissimilarity functions outperform the Averaged Bands NDD and the Spectral dissimilarity functions.
\begin{table}
\caption{\label{tab:ANR_Forests}ANR values of the hyperspectral RF-CBIR for the Forests categorical search.}
{\footnotesize }%
\begin{tabular}{|l|c|c|c|c|c|c|}
\cline{4-7} 
\multicolumn{3}{c|}{} & {\footnotesize Avg.Band NDD} & {\footnotesize By-Band NDD} & {\footnotesize Spectral} & {\footnotesize Spectral-Spatial}\tabularnewline
\hline 
\multicolumn{3}{|c|}{{\footnotesize Zero Query}} & 0.0809 & 0.0613 & 0.1360 & 0.0552 \tabularnewline
\hline 
\multirow{6}{*}{{\footnotesize Online Prot.}} & \multirow{3}{*}{{\footnotesize 7NN}} & {\footnotesize BW} & 0.0343 & 0.0426 & 0.1394 & 0.0630\tabularnewline
\cline{3-7} 
 &  & {\footnotesize AL} & \textbf{0.0280} & \textbf{0.0258} & 0.0869 & 0.0337\tabularnewline
\cline{3-7} 
 &  & {\footnotesize BW+AL} & 0.0287 & 0.0281 & \textbf{0.0770} & \textbf{0.0330}\tabularnewline
\cline{2-7} 
 & \multirow{3}{*}{{\footnotesize SVM}} & {\footnotesize BW} & 0.0383 & 0.1392 & 0.2600 & 0.0852\tabularnewline
\cline{3-7} 
 &  & {\footnotesize AL} & 0.0596 & 0.1155 & 0.3947 & 0.2371\tabularnewline
\cline{3-7} 
 &  & {\footnotesize BW+AL} & 0.0462 & 0.0358 & 0.2143 & 0.2430\tabularnewline
\hline 
\multirow{6}{*}{{\footnotesize Offline Prot.}} & \multirow{3}{*}{{\footnotesize 7NN}} & {\footnotesize BW} & 0.0662 & 0.0723 & 0.1922 & 0.0543\tabularnewline
\cline{3-7} 
 &  & {\footnotesize AL} & 0.0329 & 0.0631 & 0.1735 & 0.0494\tabularnewline
\cline{3-7} 
 &  & {\footnotesize BW+AL} & 0.0448 & 0.0633 & 0.1848 & 0.0473\tabularnewline
\cline{2-7} 
 & \multirow{3}{*}{{\footnotesize SVM}} & {\footnotesize BW} & 0.0758 & 0.0478 & 0.2502 & 0.1063\tabularnewline
\cline{3-7} 
 &  & {\footnotesize AL} & 0.0542 & 0.0409 & 0.3116 & 0.1678\tabularnewline
\cline{3-7} 
 &  & {\footnotesize BW+AL} & 0.0642 & 0.0538 & 0.3180 & 0.1055\tabularnewline
\hline 
\end{tabular}
\end{table}

\begin{table}
\caption{\label{tab:ANR_Fields}ANR values of the hyperspectral RF-CBIR for the Fields categorical search.}
{\footnotesize }%
\begin{tabular}{|l|c|c|c|c|c|c|}
\cline{4-7} 
\multicolumn{3}{c|}{} & {\footnotesize Avg.Band NDD} & {\footnotesize By-Band NDD} & {\footnotesize Spectral} & {\footnotesize Spectral-Spatial}\tabularnewline
\hline 
\multicolumn{3}{|c|}{{\footnotesize Zero Query}} & 0.2171 & 0.1641 & 0.1594 & 0.1599\tabularnewline
\hline 
\multirow{6}{*}{{\footnotesize Online Prot.}} & \multirow{3}{*}{{\footnotesize 7NN}} & {\footnotesize BW} & 0.1552 & 0.0634 & 0.1776 & 0.1494\tabularnewline
\cline{3-7} 
 &  & {\footnotesize AL} & \textbf{0.1388} & \textbf{0.0495} & 0.1573 & 0.1514\tabularnewline
\cline{3-7} 
 &  & {\footnotesize BW+AL} & 0.1433 & 0.0587 & 0.1862 & 0.1883\tabularnewline
\cline{2-7} 
 & \multirow{3}{*}{{\footnotesize SVM}} & {\footnotesize BW} & 0.1898 & 0.2462 & 0.1511 & 0.1983\tabularnewline
\cline{3-7} 
 &  & {\footnotesize AL} & 0.1808 & 0.0914 & 0.1526 & \textbf{0.0924}\tabularnewline
\cline{3-7} 
 &  & {\footnotesize BW+AL} & 0.1567 & 0.0812 & \textbf{0.1477} & 0.1184\tabularnewline
\hline 
\multirow{6}{*}{{\footnotesize Offline Prot.}} & \multirow{3}{*}{{\footnotesize 7NN}} & {\footnotesize BW} & 0.1847 & 0.0756 & 0.2607 & 0.1779\tabularnewline
\cline{3-7} 
 &  & {\footnotesize AL} & 0.1802 & 0.0533 & 0.2660 & 0.2158\tabularnewline
\cline{3-7} 
 &  & {\footnotesize BW+AL} & 0.1694 & 0.0569 & 0.2957 & 0.1994\tabularnewline
\cline{2-7} 
 & \multirow{3}{*}{{\footnotesize SVM}} & {\footnotesize BW} & 0.2033& 0.0724 & 0.2136 & 0.1936\tabularnewline
\cline{3-7} 
 &  & {\footnotesize AL} & 0.1831 & 0.0660 & 0.2112 & 0.1700\tabularnewline
\cline{3-7} 
 &  & {\footnotesize BW+AL} & 0.2008 & 0.0497 & 0.2171 & 0.1442\tabularnewline
\hline 
\end{tabular}
\end{table}

\begin{table}
\caption{\label{tab:ANR_Urbanareas}ANR values of the hyperspectral RF-CBIR for the Urban areas categorical search.}
{\footnotesize }%
\begin{tabular}{|l|c|c|c|c|c|c|}
\cline{4-7} 
\multicolumn{3}{c|}{} & {\footnotesize Avg.Band NDD} & {\footnotesize By-Band NDD} & {\footnotesize Spectral} & {\footnotesize Spectral-Spatial}\tabularnewline
\hline 
\multicolumn{3}{|c|}{{\footnotesize Zero Query}} & \textbf{0.1217} & \textbf{0.0080} & 0.2068 & 0.0732\tabularnewline
\hline 
\multirow{6}{*}{{\footnotesize Online Prot.}} & \multirow{3}{*}{{\footnotesize 7NN}} & {\footnotesize BW} & 0.1920 & \textbf{\textit{0.0082}} & \textbf{0.0509} & 0.0416\tabularnewline
\cline{3-7} 
 &  & {\footnotesize AL} & 0.1900 & 0.0096 & 0.0626 & \textbf{0.0392}\tabularnewline
\cline{3-7} 
 &  & {\footnotesize BW+AL} & 0.2702 & 0.0282 & 0.1230 & 0.0654\tabularnewline
\cline{2-7} 
 & \multirow{3}{*}{{\footnotesize SVM}} & {\footnotesize BW} & 0.2675 & 0.0437 & 0.1120 & 0.2126\tabularnewline
\cline{3-7} 
 &  & {\footnotesize AL} & 0.5870 & 0.0416 & 0.2501 & 0.1603\tabularnewline
\cline{3-7} 
 &  & {\footnotesize BW+AL} & 0.3825 & 0.0415 & 0.1459 & 0.1712\tabularnewline
\hline 
\multirow{6}{*}{{\footnotesize Offline Prot.}} & \multirow{3}{*}{{\footnotesize 7NN}} & {\footnotesize BW} & 0.2578 & 0.0545 & 0.0799 & 0.0762\tabularnewline
\cline{3-7} 
 &  & {\footnotesize AL} & 0.2713 & 0.0276 & 0.0698 & 0.0570\tabularnewline
\cline{3-7} 
 &  & {\footnotesize BW+AL} & 0.3425 & 0.1061 & 0.1509 & 0.1224\tabularnewline
\cline{2-7} 
 & \multirow{3}{*}{{\footnotesize SVM}} & {\footnotesize BW} & \textbf{\textit{0.1562}} & 0.0103 & 0.0833 & 0.1240\tabularnewline
\cline{3-7} 
 &  & {\footnotesize AL} & 0.2276 & 0.0273 & 0.2164 & 0.2651\tabularnewline
\cline{3-7} 
 &  & {\footnotesize BW+AL} & 0.1763 & 0.0246 & 0.0561 & 0.2032\tabularnewline
\hline 
\end{tabular}
\end{table}

\begin{figure}
\begin{tabular}{ccc}
 BW & AL \tabularnewline
 \includegraphics[width=0.4\columnwidth]{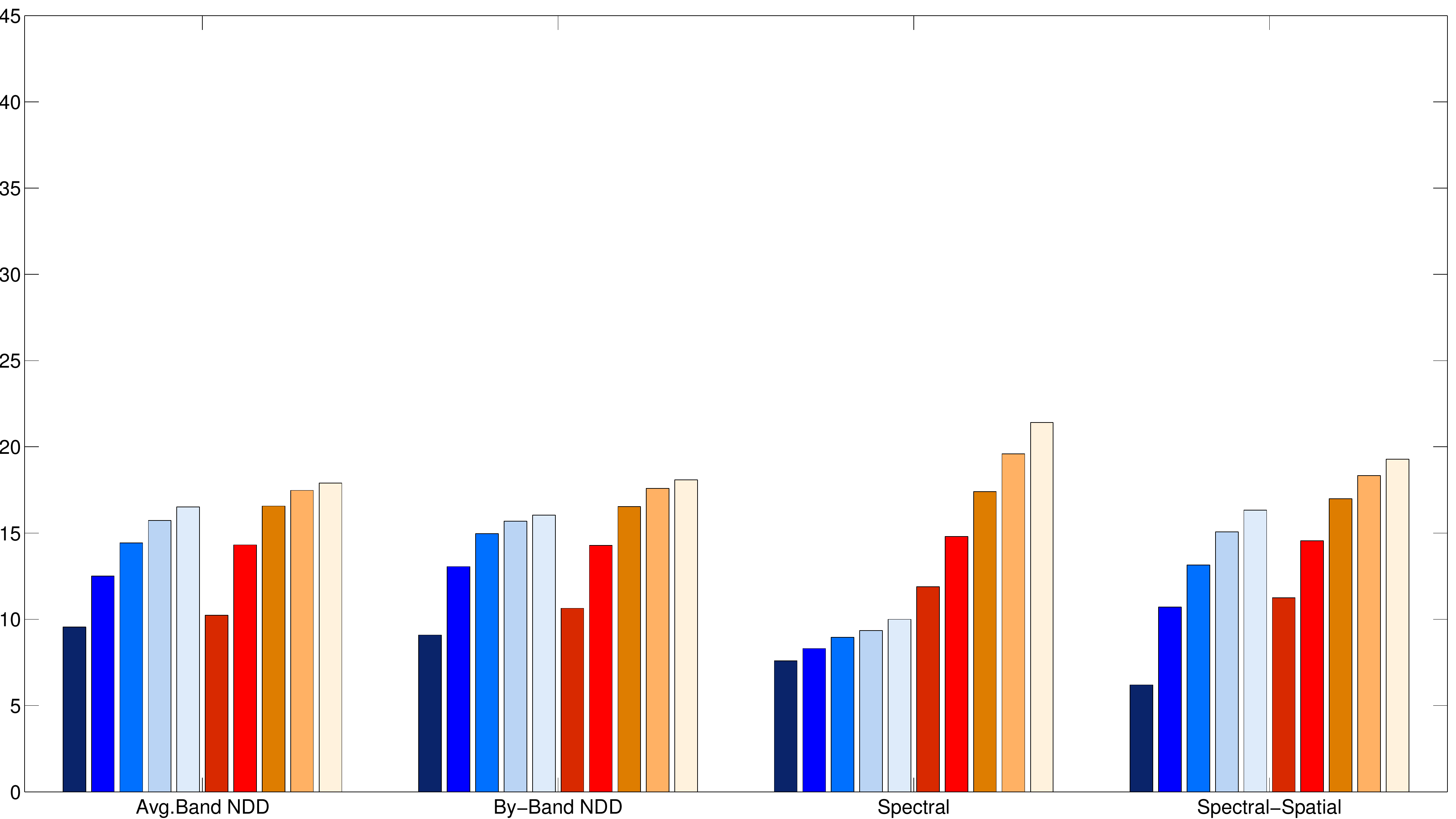} 
 & \includegraphics[width=0.4\columnwidth]{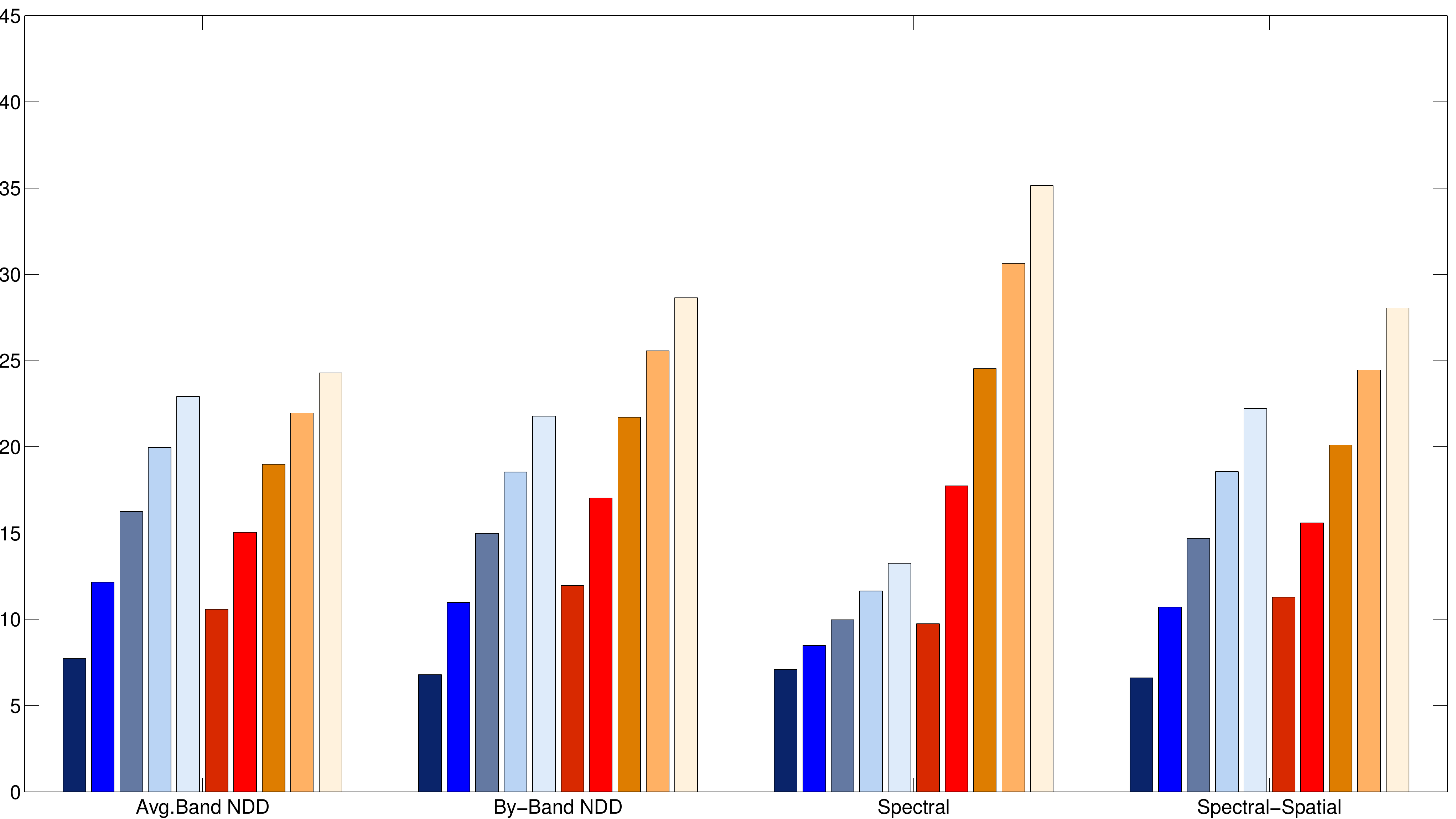} 
& \multirow{4}{*}{\includegraphics[width=0.15\columnwidth]{ts_legend}}
\tabularnewline
\multicolumn{2}{c}{Online Prototypes - 7NN} &
\tabularnewline
 \includegraphics[width=0.4\columnwidth]{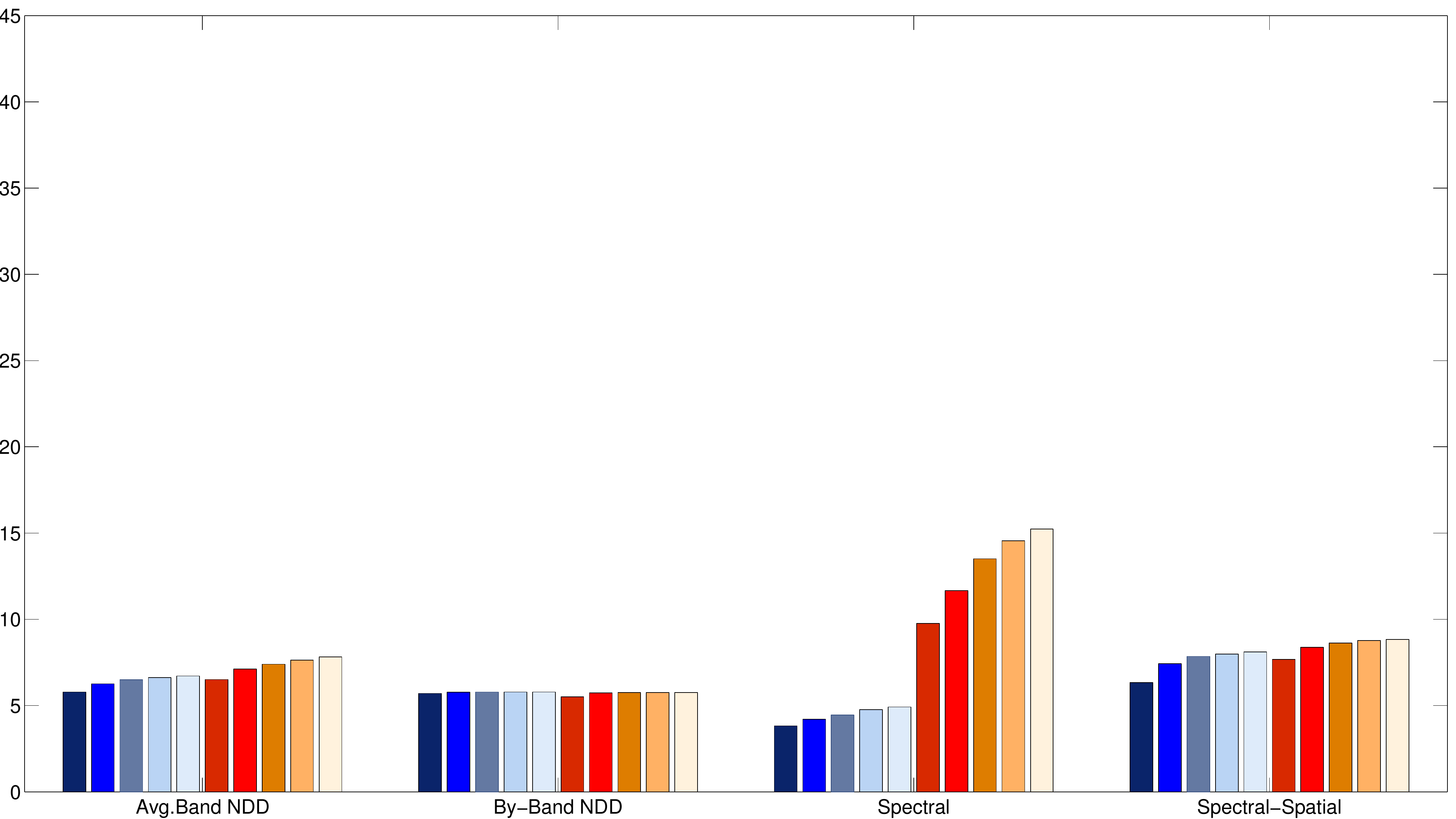} 
 & \includegraphics[width=0.4\columnwidth]{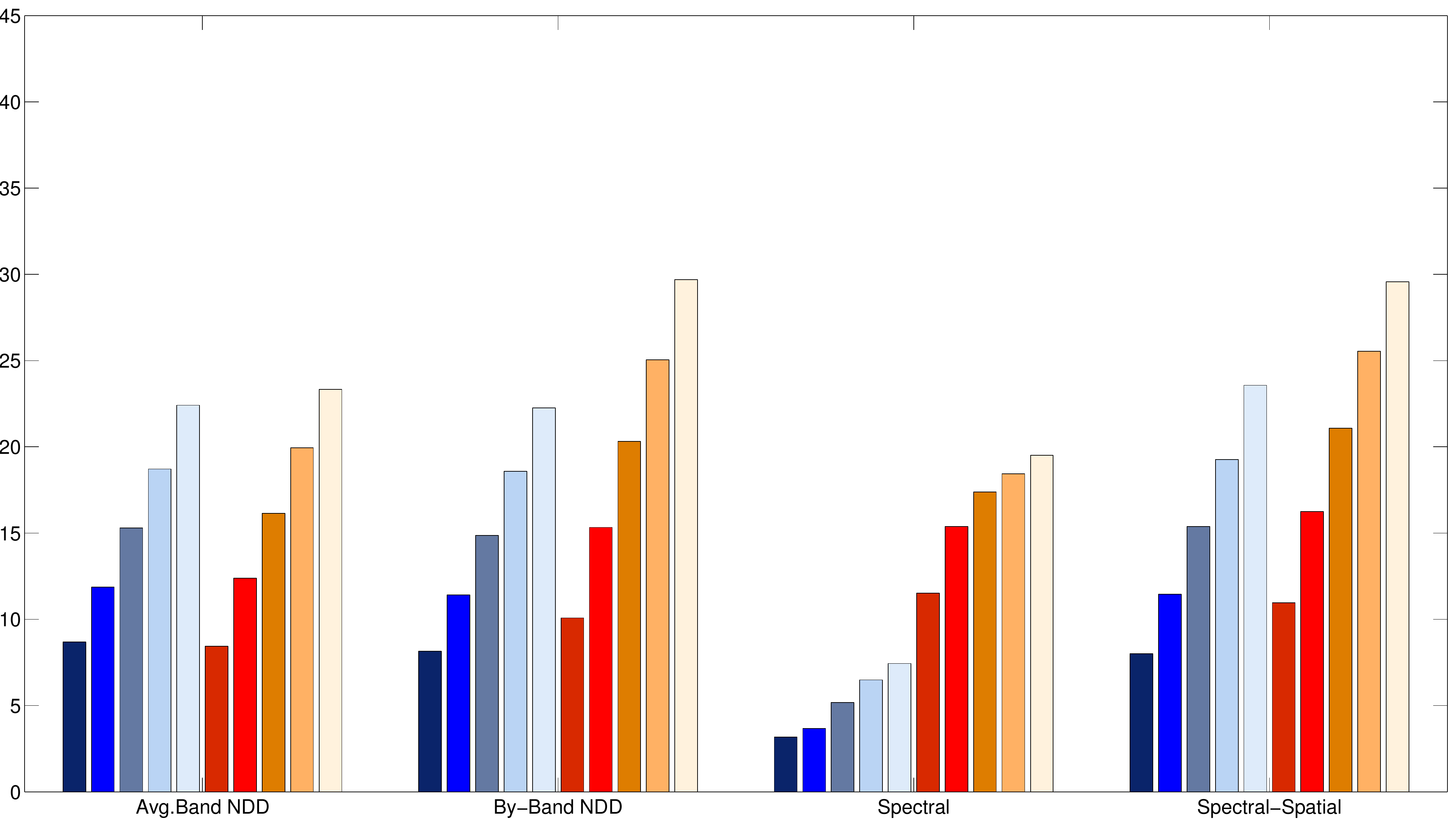} 
& \tabularnewline
\multicolumn{2}{c}{Online Prototypes - SVM} &
\tabularnewline
 \includegraphics[width=0.4\columnwidth]{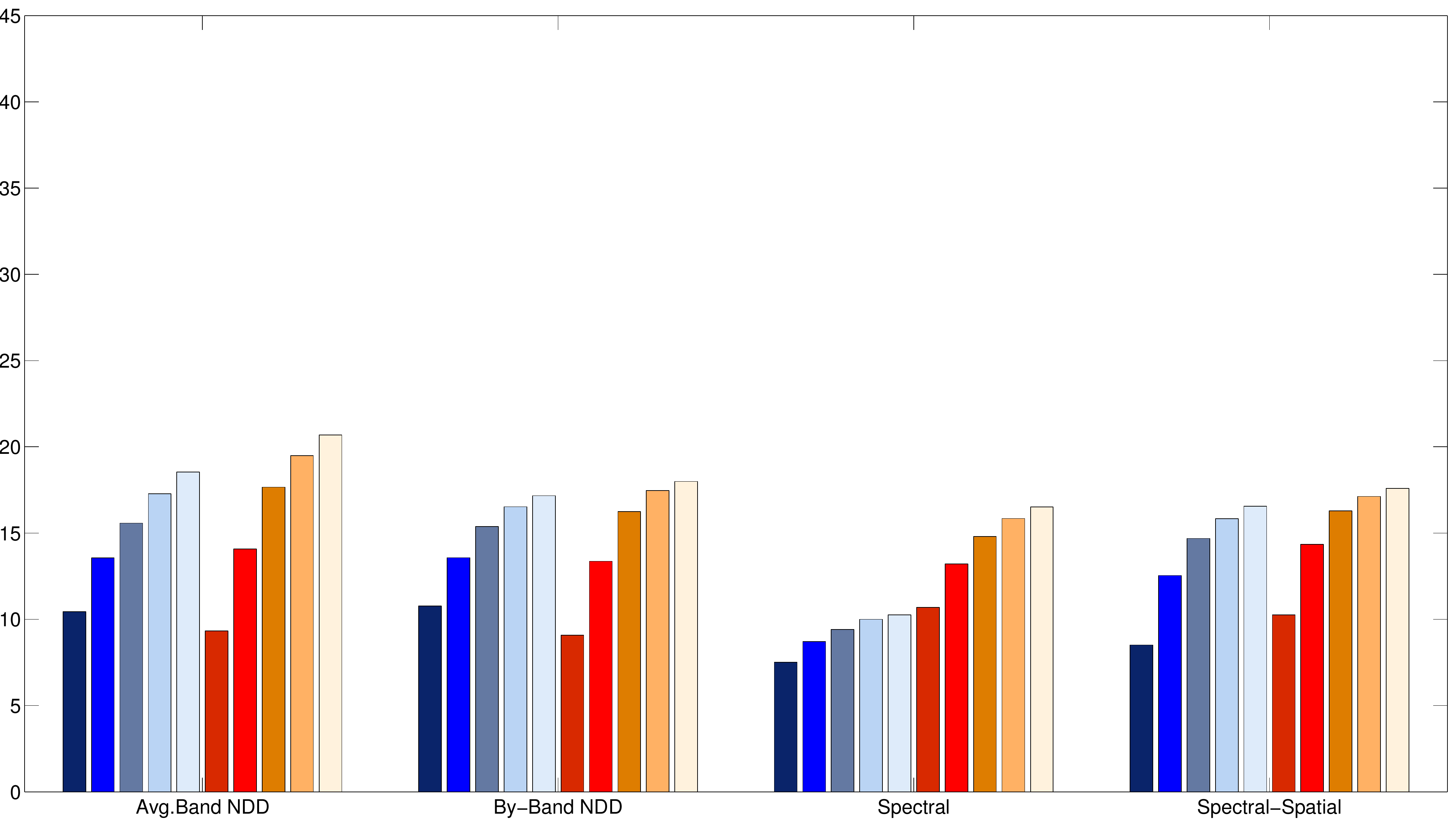} 
 & \includegraphics[width=0.4\columnwidth]{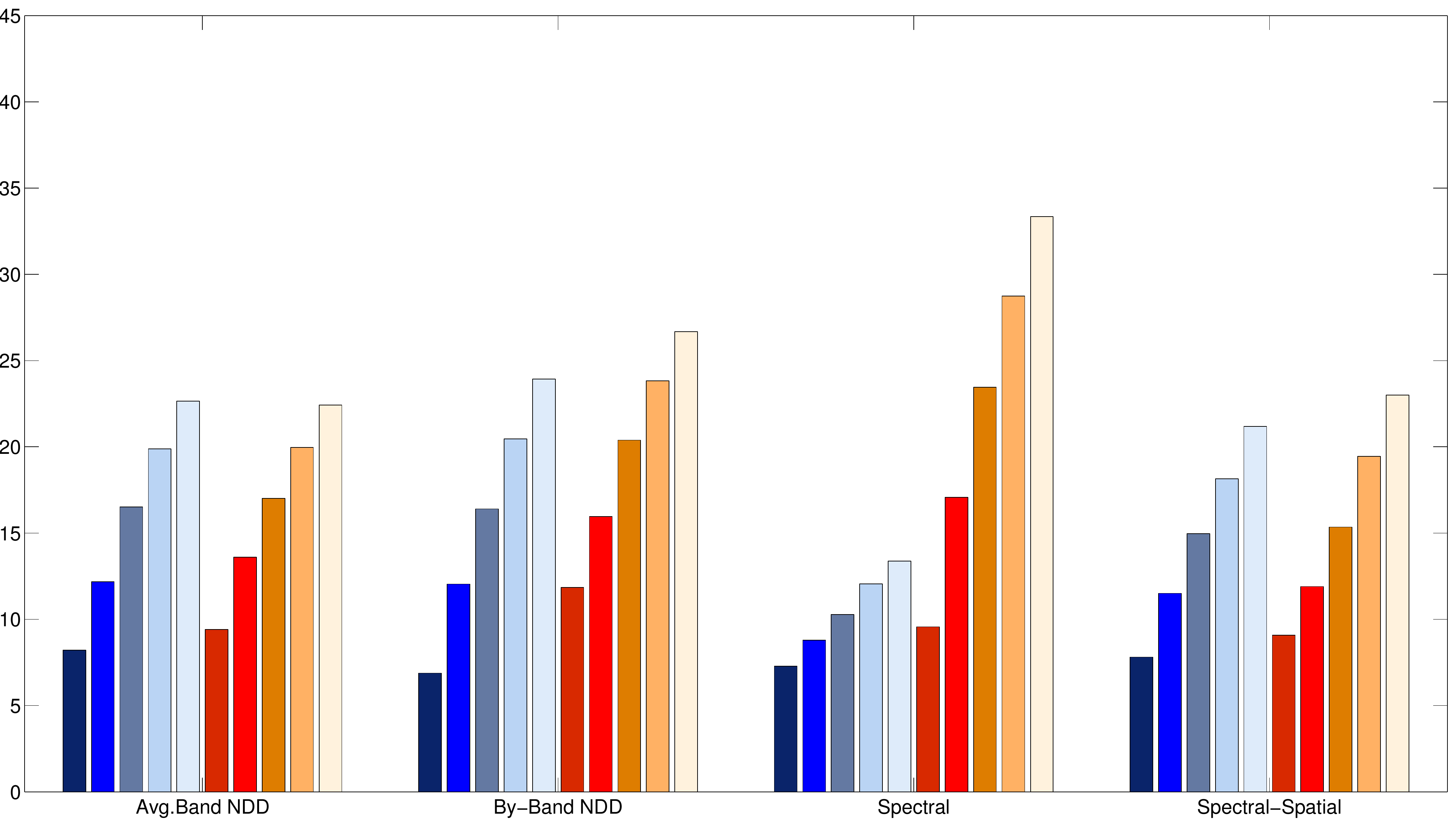} 
& \tabularnewline
\multicolumn{2}{c}{Offline Prototypes - 7NN} &
\tabularnewline
 \includegraphics[width=0.4\columnwidth]{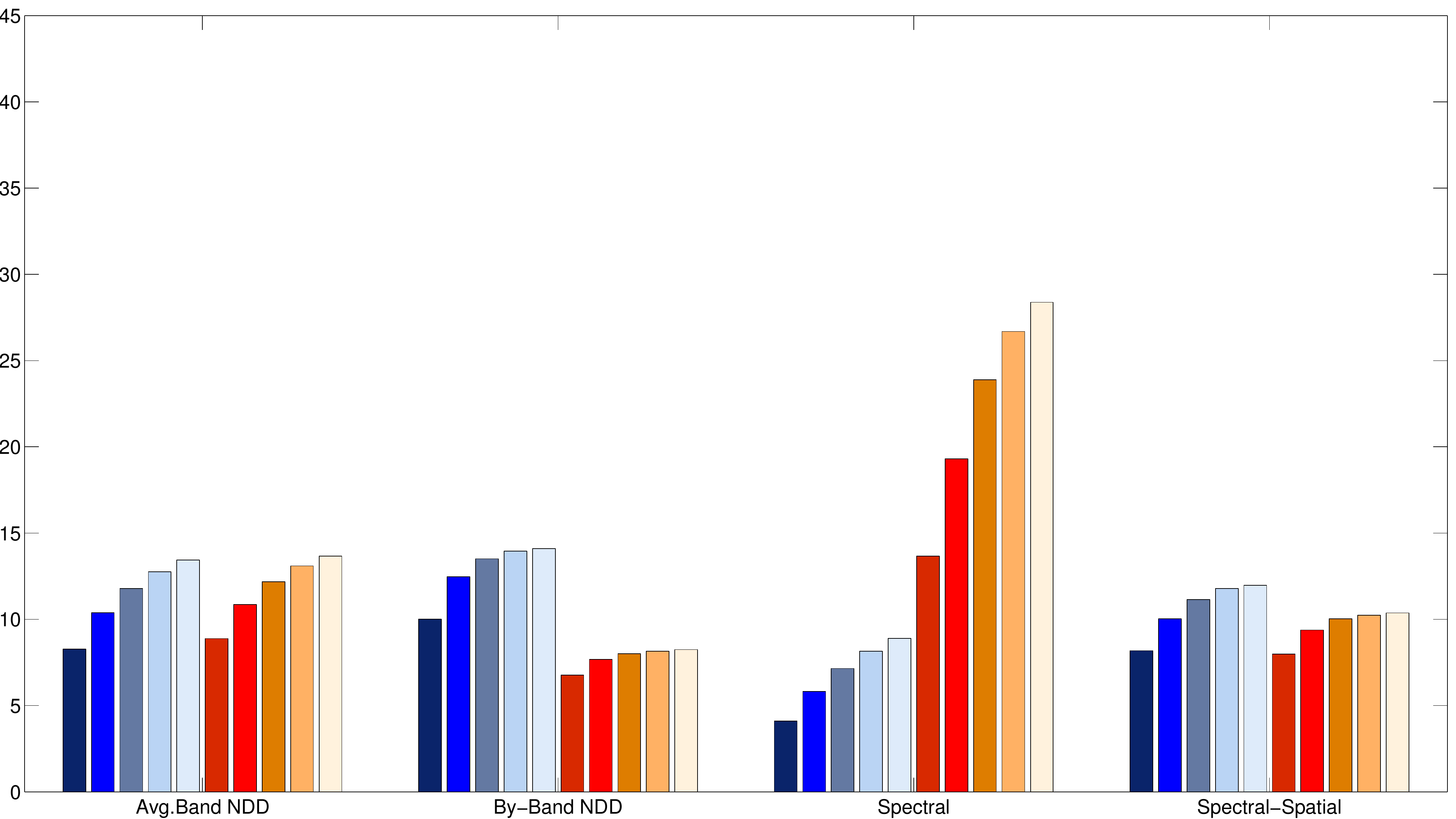} 
 & \includegraphics[width=0.4\columnwidth]{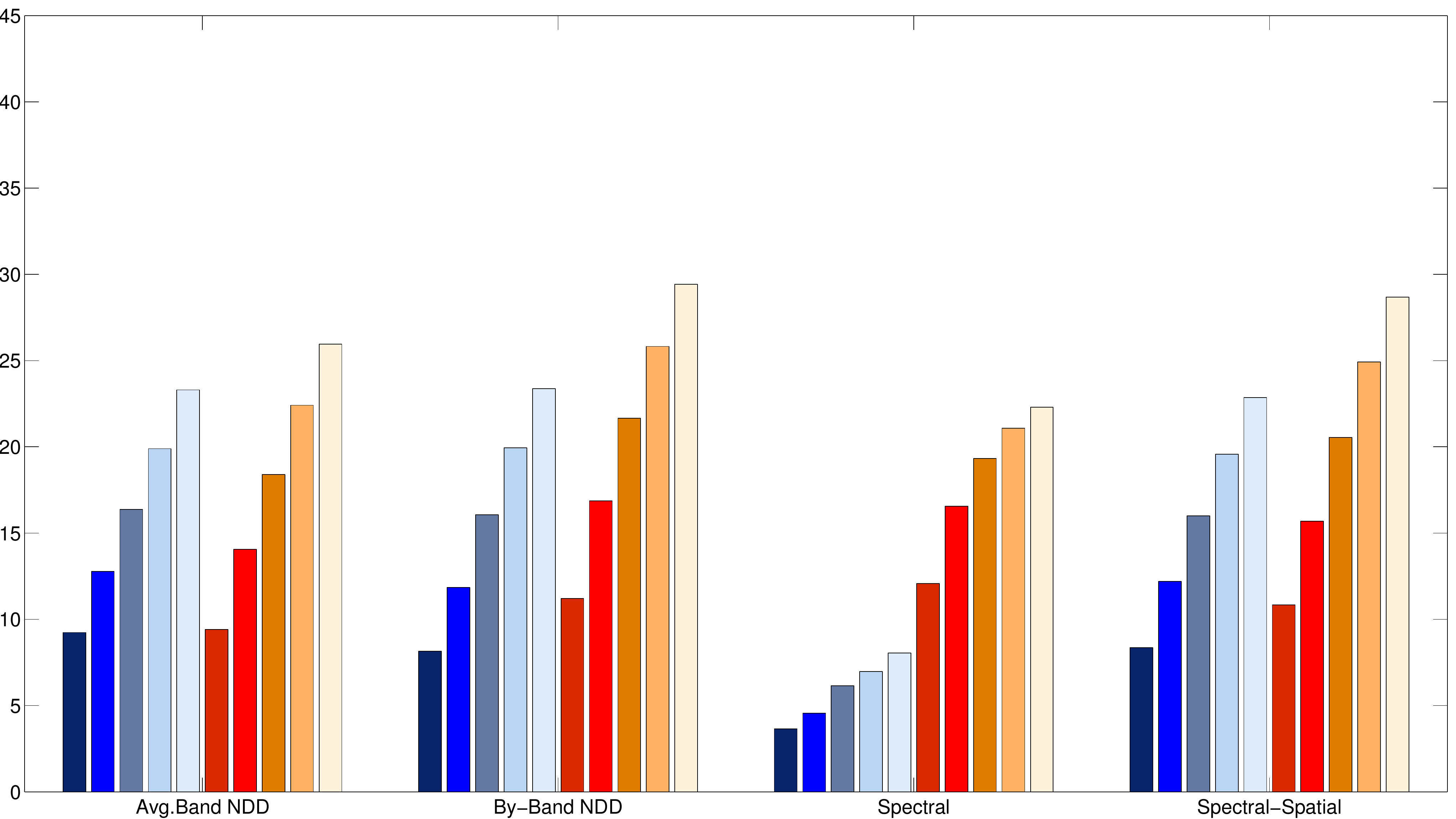} 
& \tabularnewline
\multicolumn{2}{c}{Offline Prototypes - SVM} &
\tabularnewline
\end{tabular}
\caption{\label{fig:TS_Fields}Average number of relevant (R) and non-relevant (NR) images in the training set for each RF iteration and comparing hyperspectral dissimilarity functions, using the BW and the AL image retrieval selection criteria for the Fields categorical search.}
\end{figure}

\begin{figure}
\begin{tabular}{ccc}
 BW & AL \tabularnewline
 \includegraphics[width=0.4\columnwidth]{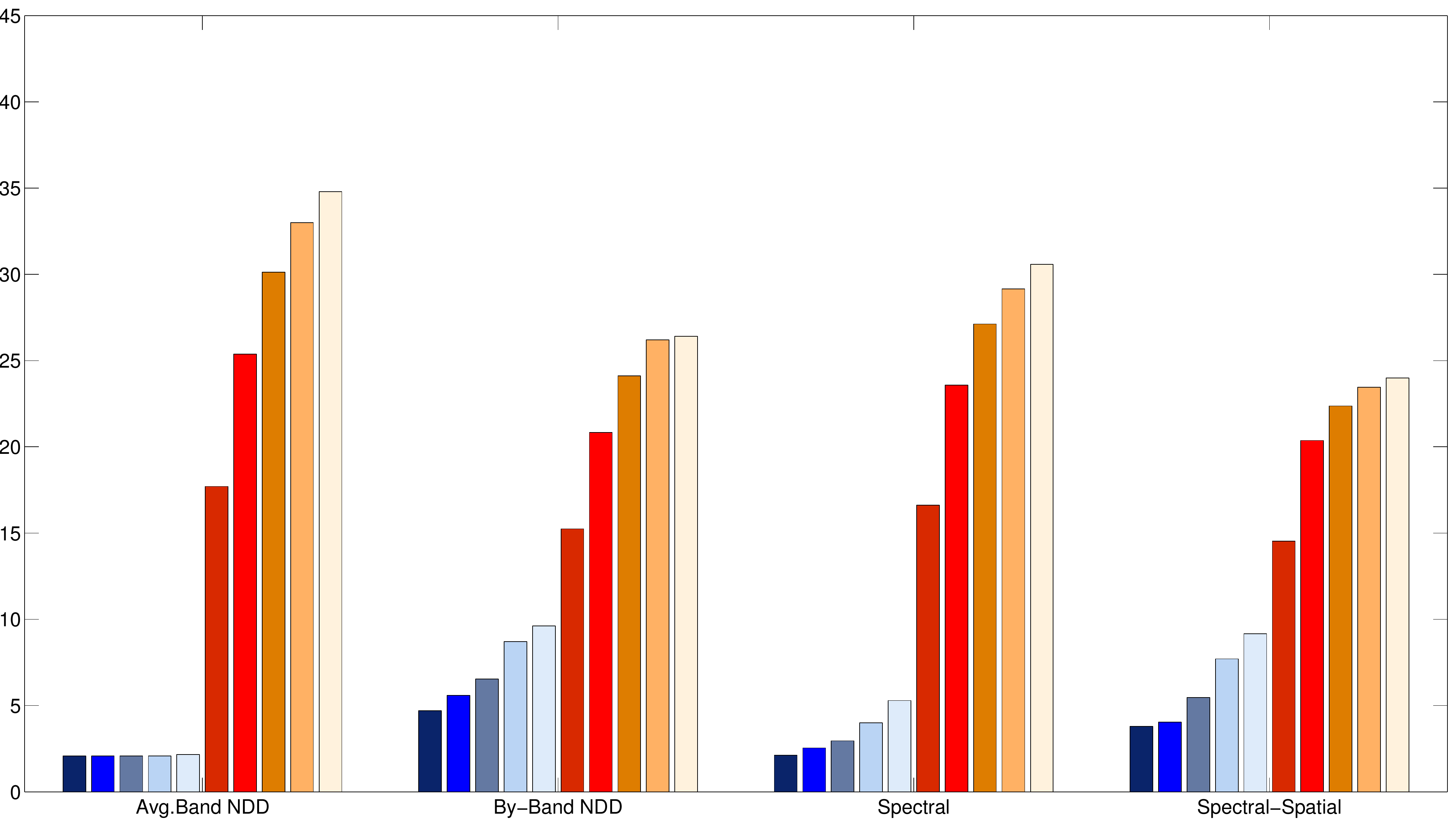} 
 & \includegraphics[width=0.4\columnwidth]{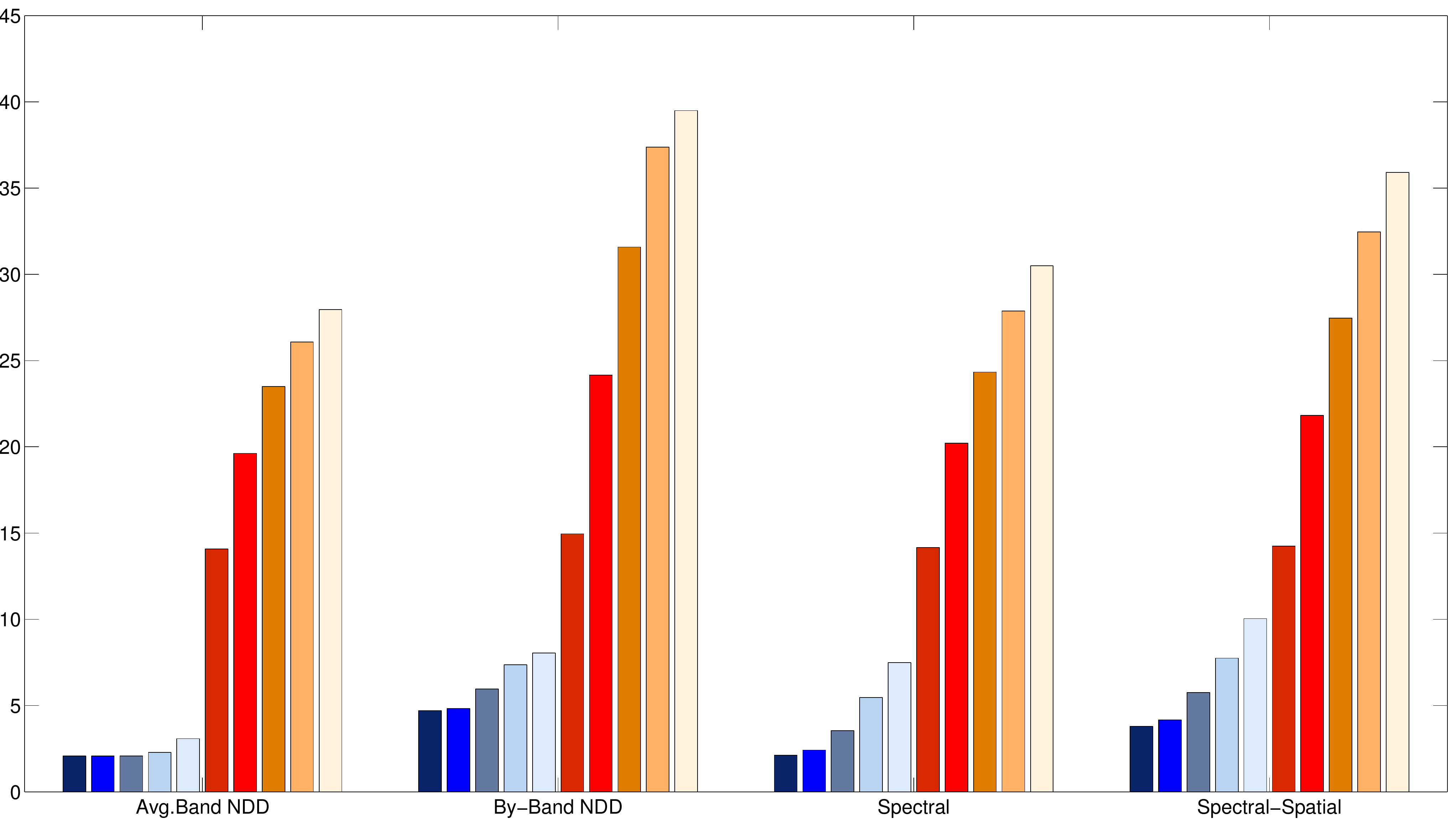} 
& \multirow{4}{*}{\includegraphics[width=0.15\columnwidth]{ts_legend}}
\tabularnewline
\multicolumn{2}{c}{Online Prototypes - 7NN} &
\tabularnewline
 \includegraphics[width=0.4\columnwidth]{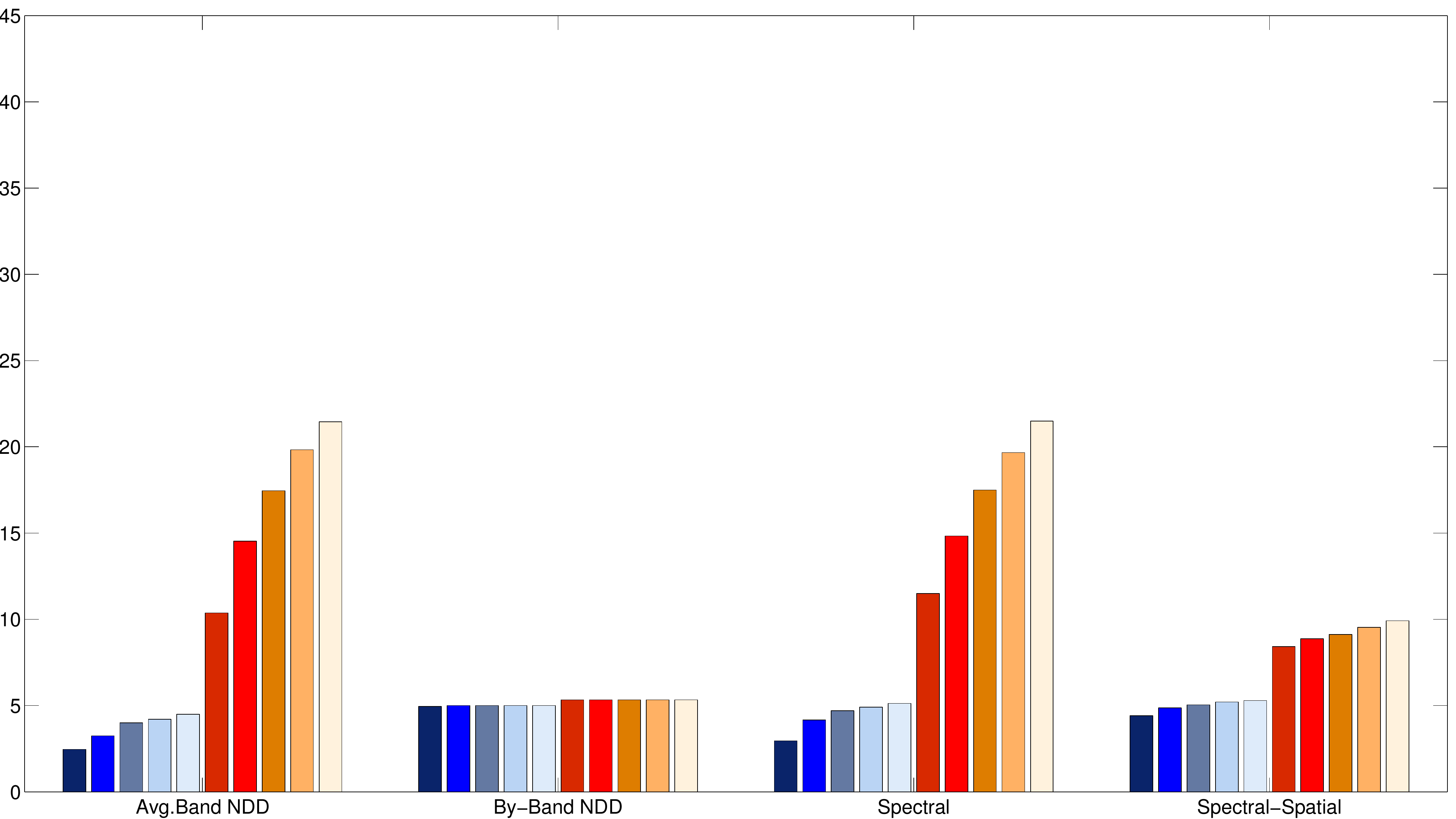} 
 & \includegraphics[width=0.4\columnwidth]{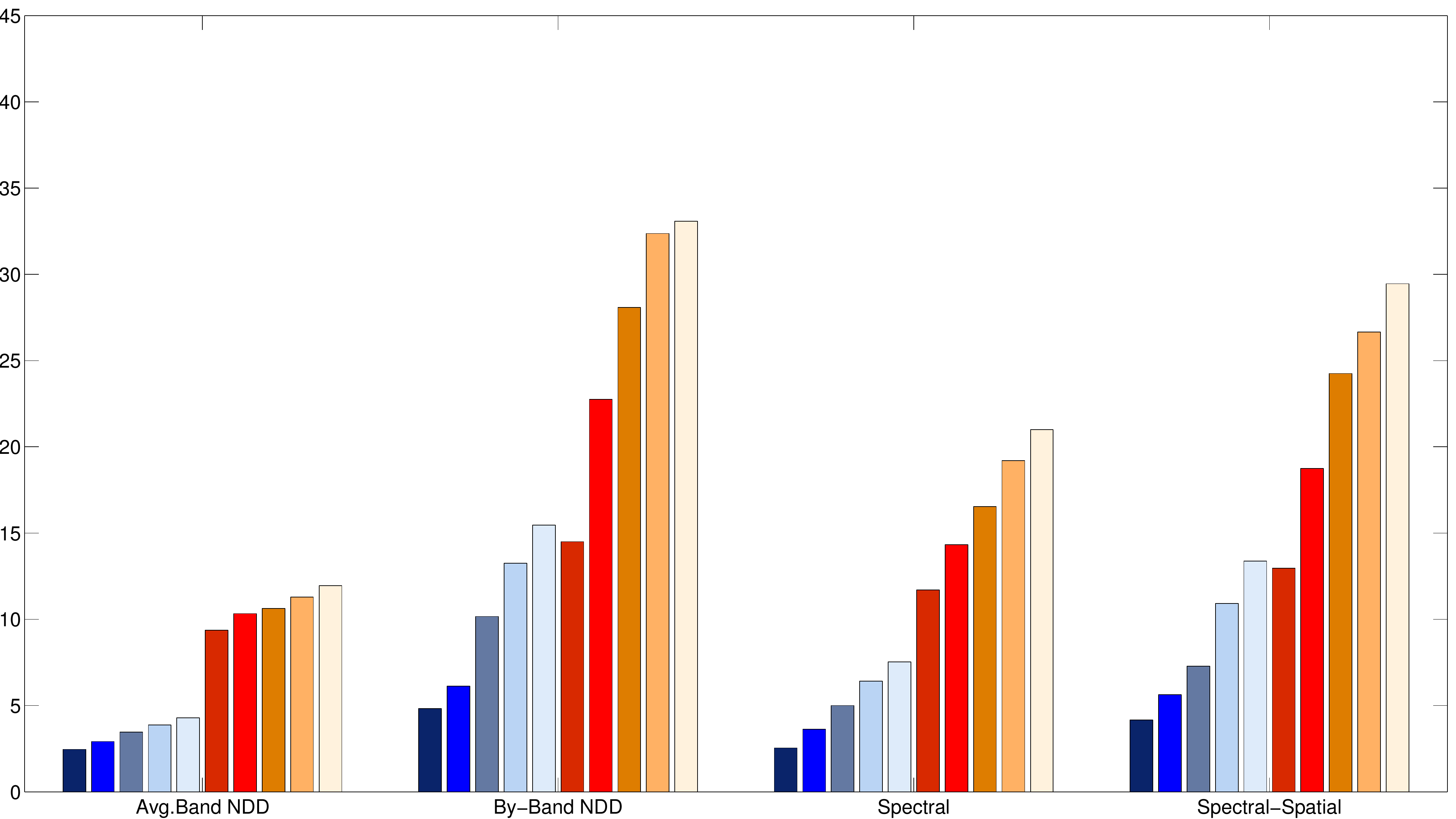} 
& \tabularnewline
\multicolumn{2}{c}{Online Prototypes - SVM} &
\tabularnewline
 \includegraphics[width=0.4\columnwidth]{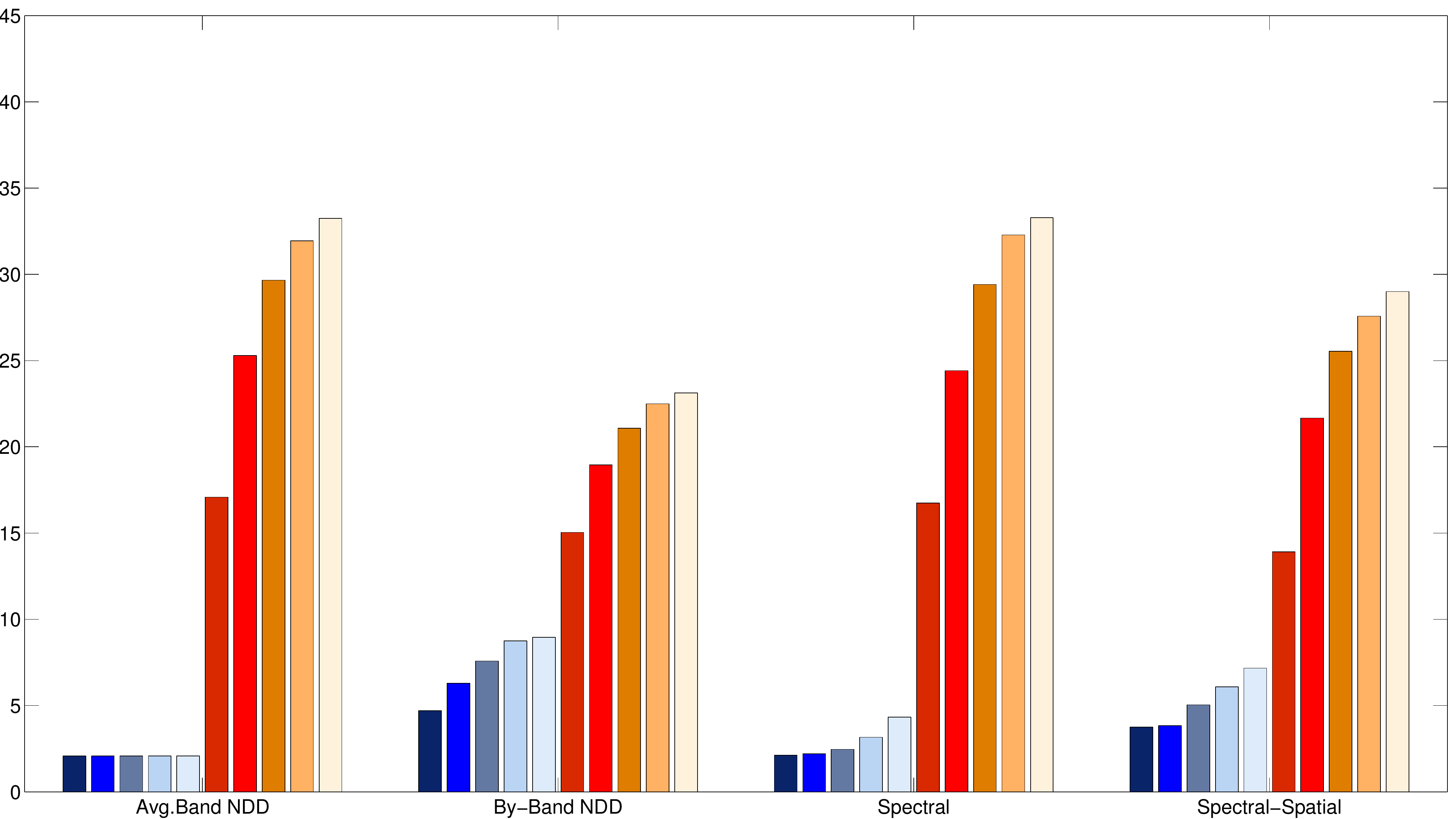} 
 & \includegraphics[width=0.4\columnwidth]{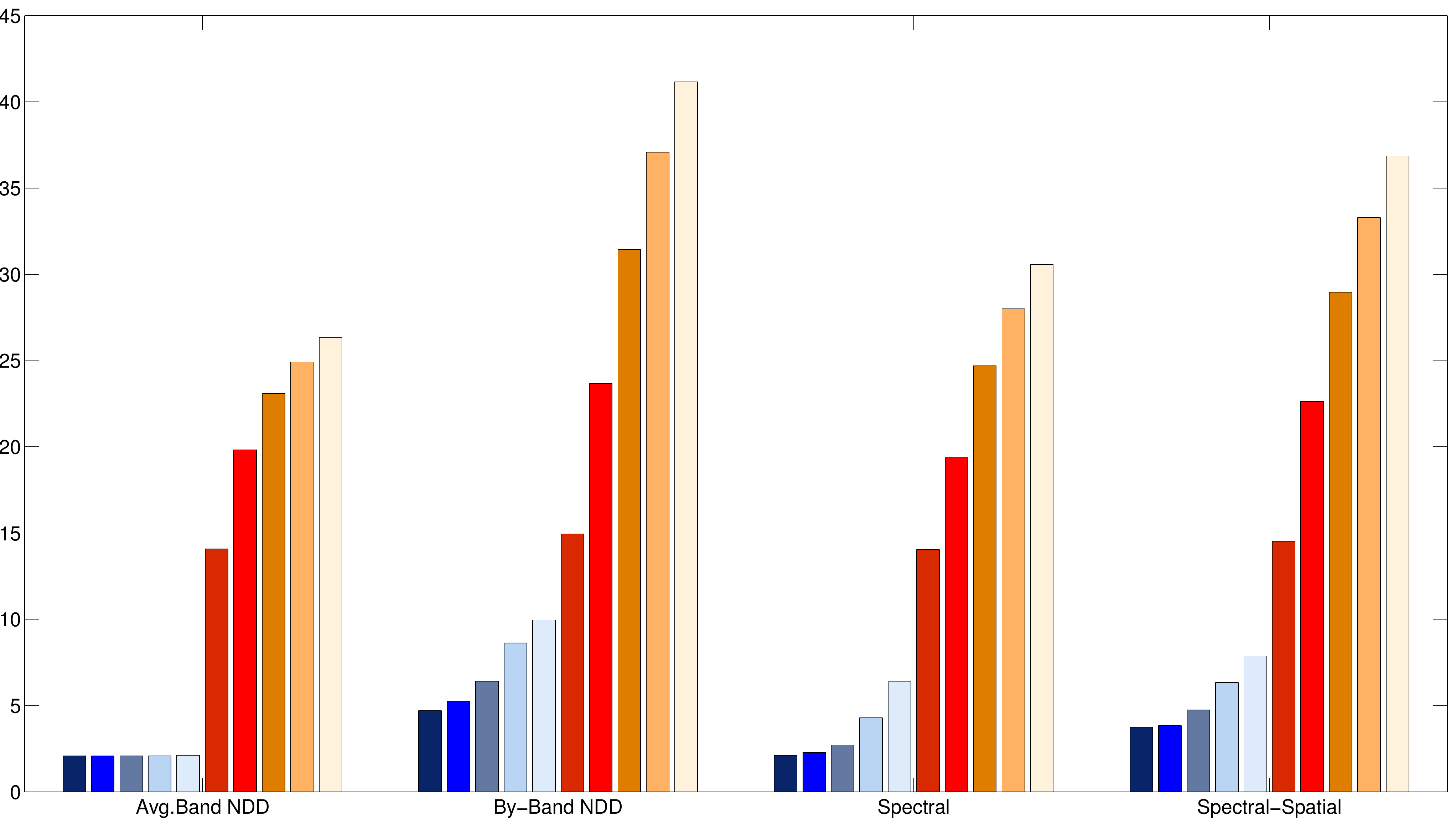} 
& \tabularnewline
\multicolumn{2}{c}{Offline Prototypes - 7NN} &
\tabularnewline
 \includegraphics[width=0.4\columnwidth]{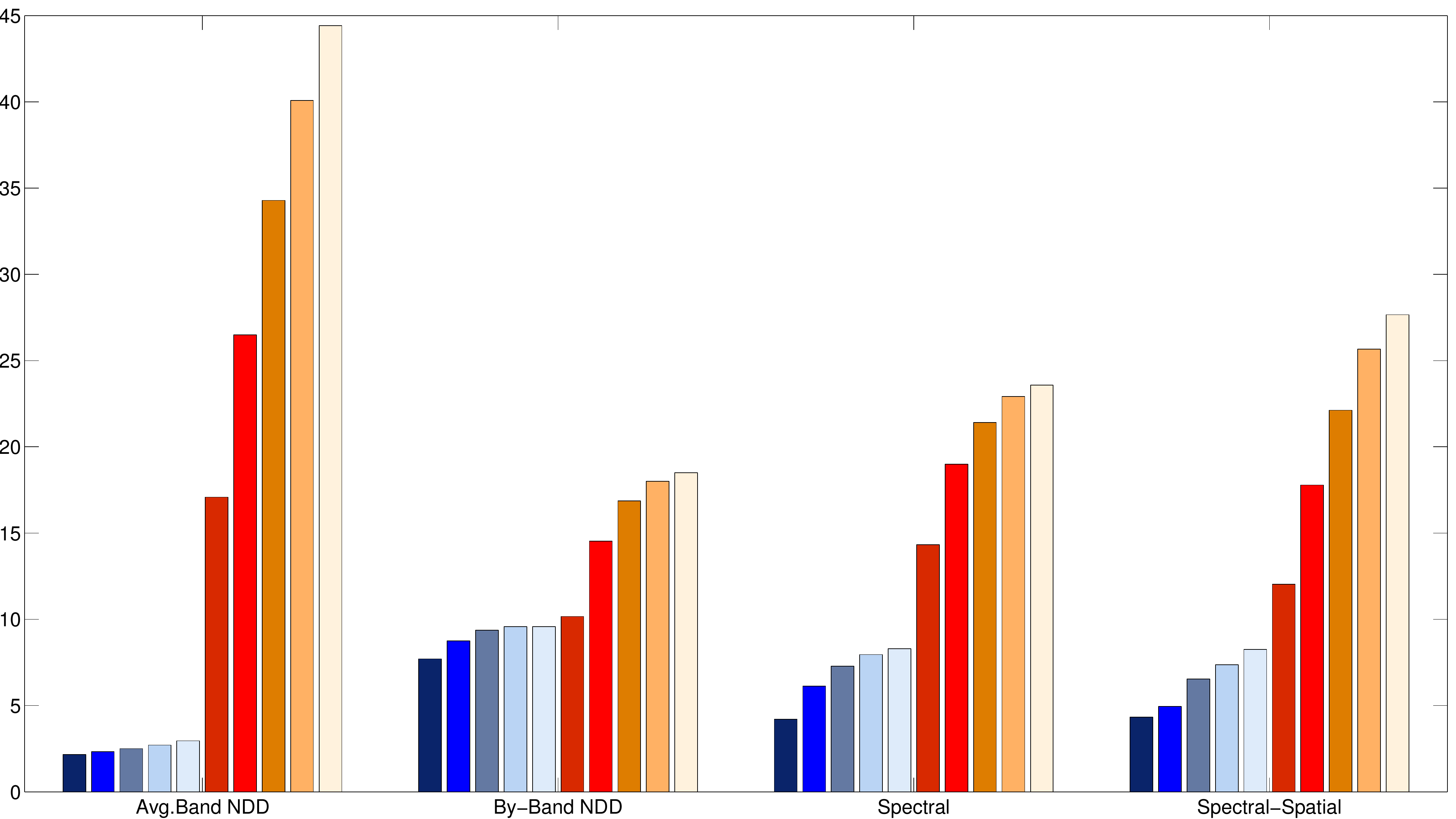} 
 & \includegraphics[width=0.4\columnwidth]{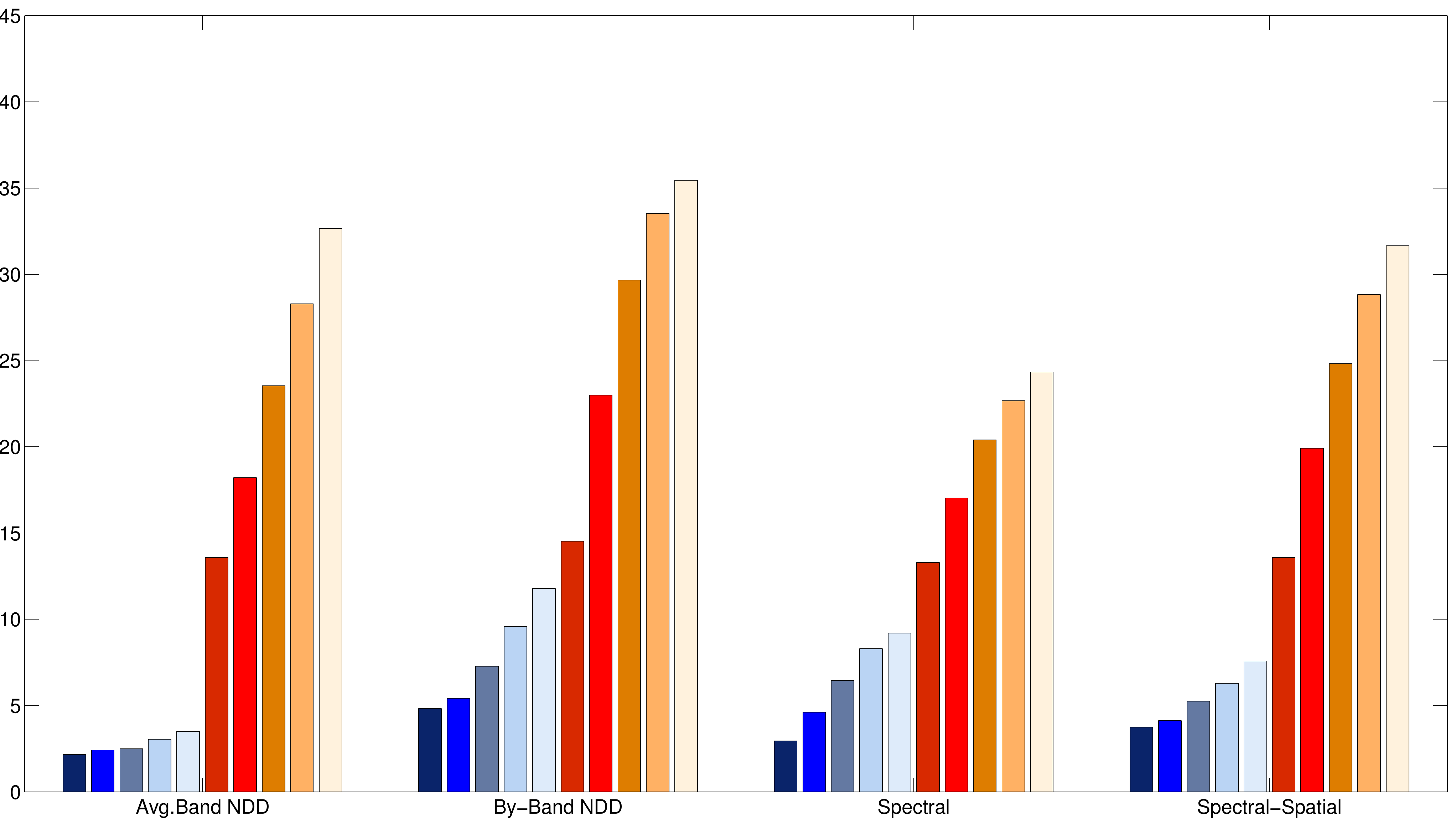} 
& \tabularnewline
\multicolumn{2}{c}{Offline Prototypes - SVM} &
\tabularnewline
\end{tabular}
\caption{\label{fig:TS_Urbans}Average number of relevant (R) and non-relevant (NR) images in the training set for each RF iteration and comparing hyperspectral dissimilarity functions, using the BW and the AL image retrieval selection criteria for the Urban areas categorical search.}
\end{figure}

There are however some discrepancies depending on the categorical query. This effect is specially relevant for the Urban areas category and it is related to the asymmetry in the number of images present in the database for each class. The low number of images belonging to the Urban areas category makes the training set very unbalanced yielding to poor classification results, and so, to a low performance in the CBIR ranking. Figures \ref{fig:TS_Fields}-\ref{fig:TS_Urbans} show the average number of relevant (R) and non-relevant (NR) images in the training set for each RF iteration using the BW and the AL image retrieval selection criteria for the Forests, Fields and Urban areas categorical queries respectively. It is clear that the Urban areas category presents the most asymmetrical distribution of the training set into relevant and non-relevant images, what it can explain the poor results on the RF process for this category. In general, the asymmetry in the R/NR ratio is not so important as soon as there are some critical number of each on the training set. It is also possible to observe that the AL selection criterion yields to better training sets compared to the BW selection criterion, expressed as bigger and more equally distributed training sets. This issue seems to be a major drawback for the SVM classifier while the mpact on the $7$-NN classifier is less severe as soon as there are enough positive samples present on the training set. This issue should be further addressed in future research in order to develop an operative hyperspectral RF-CBIR system.
\begin{figure}
\begin{tabular}{ccc}
 BW & AL & \tabularnewline
 \includegraphics[width=0.4\columnwidth]{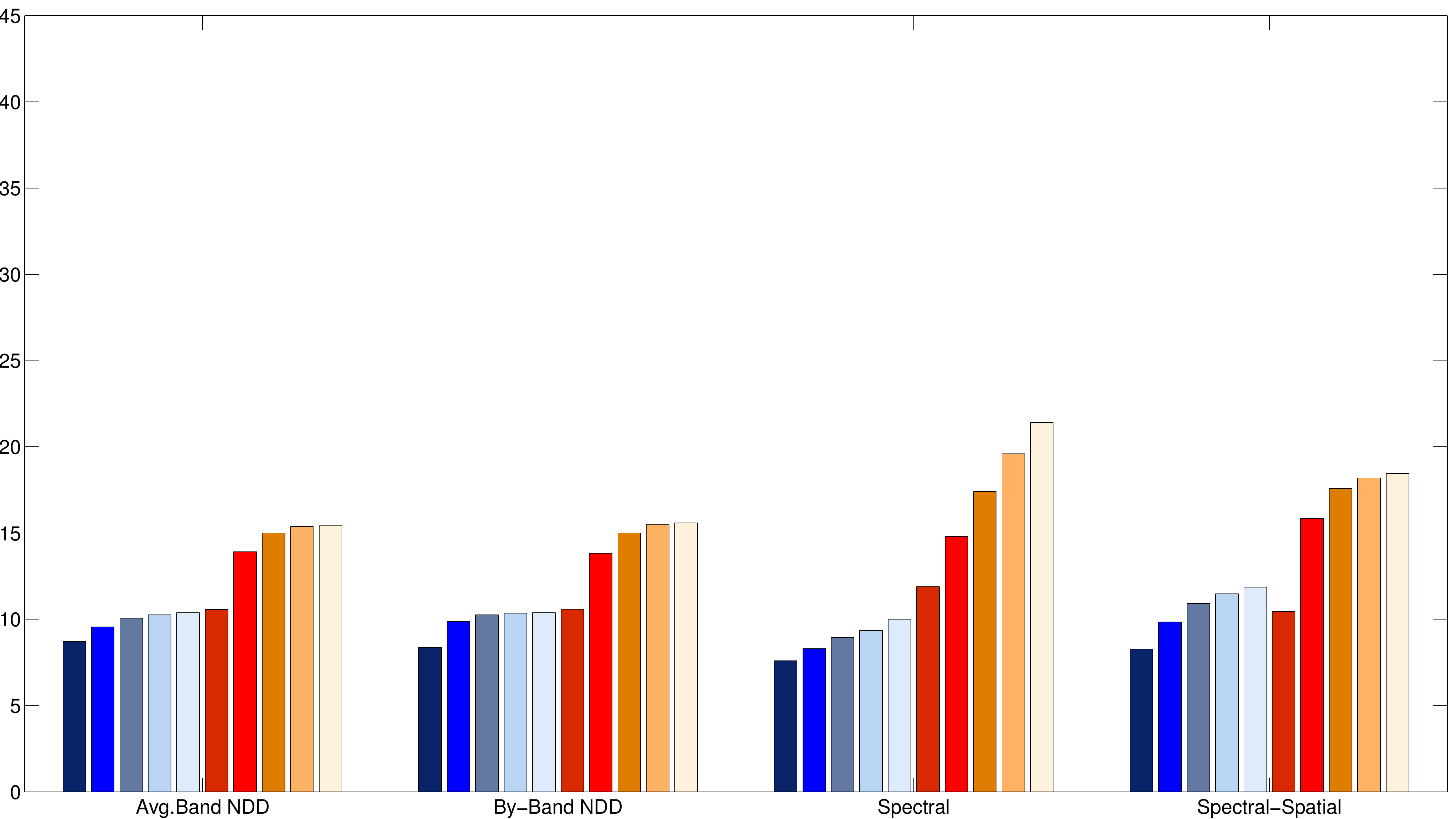} 
 & \includegraphics[width=0.4\columnwidth]{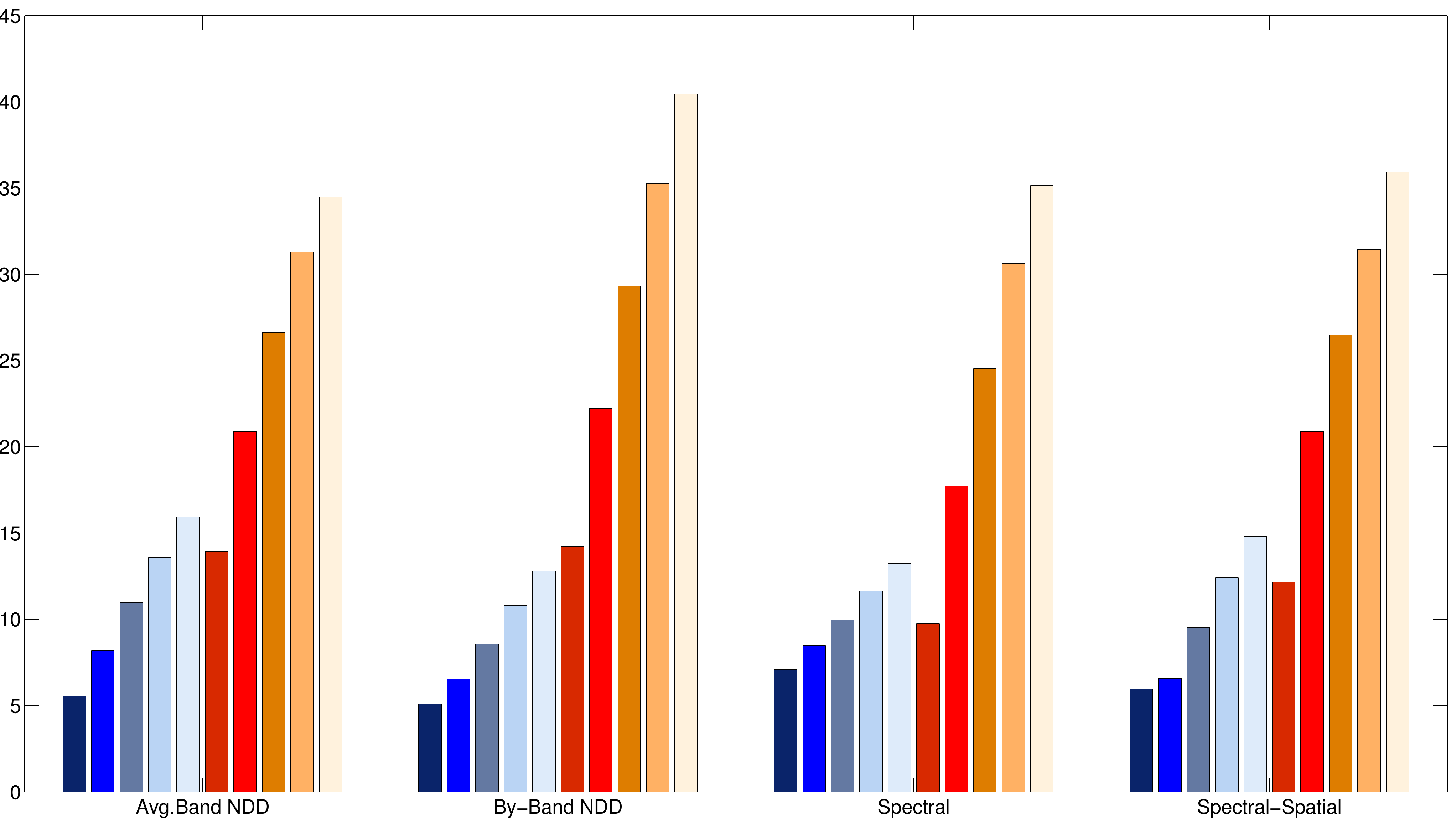} 
& \multirow{4}{*}{\includegraphics[width=0.15\columnwidth]{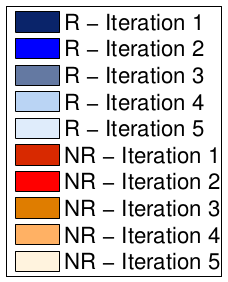}}
\tabularnewline
\multicolumn{2}{c}{Online Prototypes - 7NN} &
\tabularnewline
 \includegraphics[width=0.4\columnwidth]{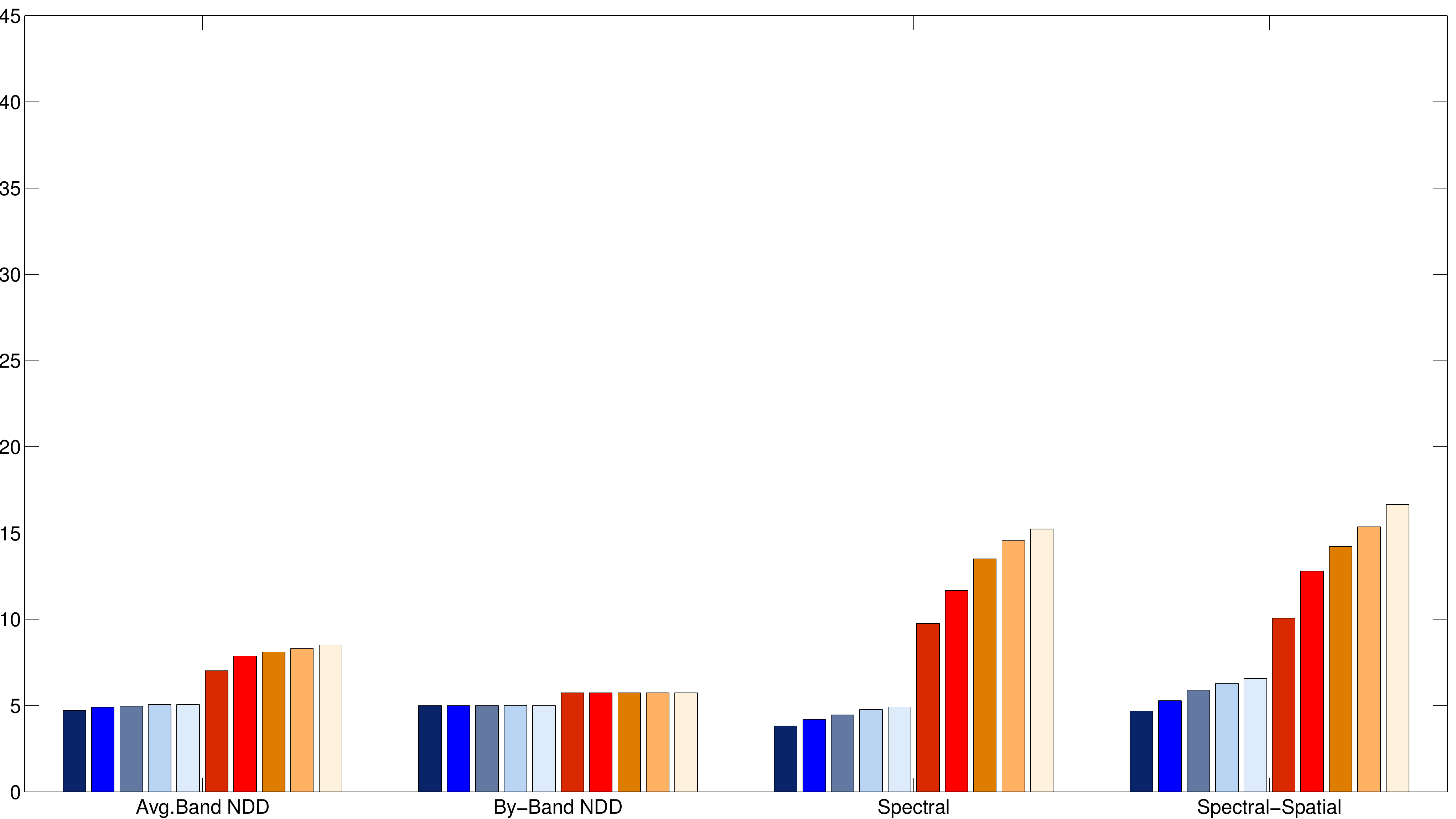} 
 & \includegraphics[width=0.4\columnwidth]{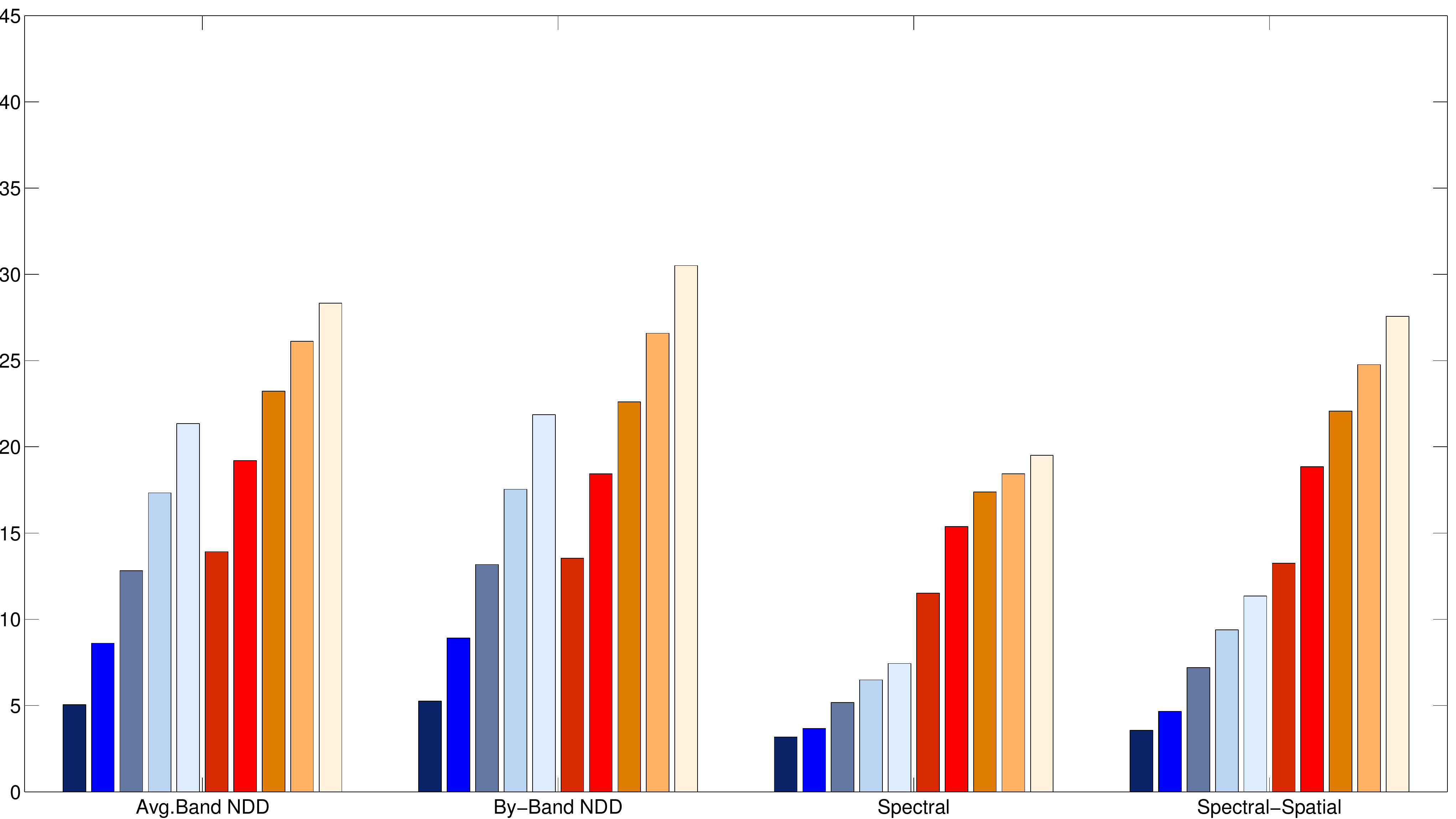} 
 & \tabularnewline
\multicolumn{2}{c}{Online Prototypes - SVM} &
\tabularnewline
 \includegraphics[width=0.4\columnwidth]{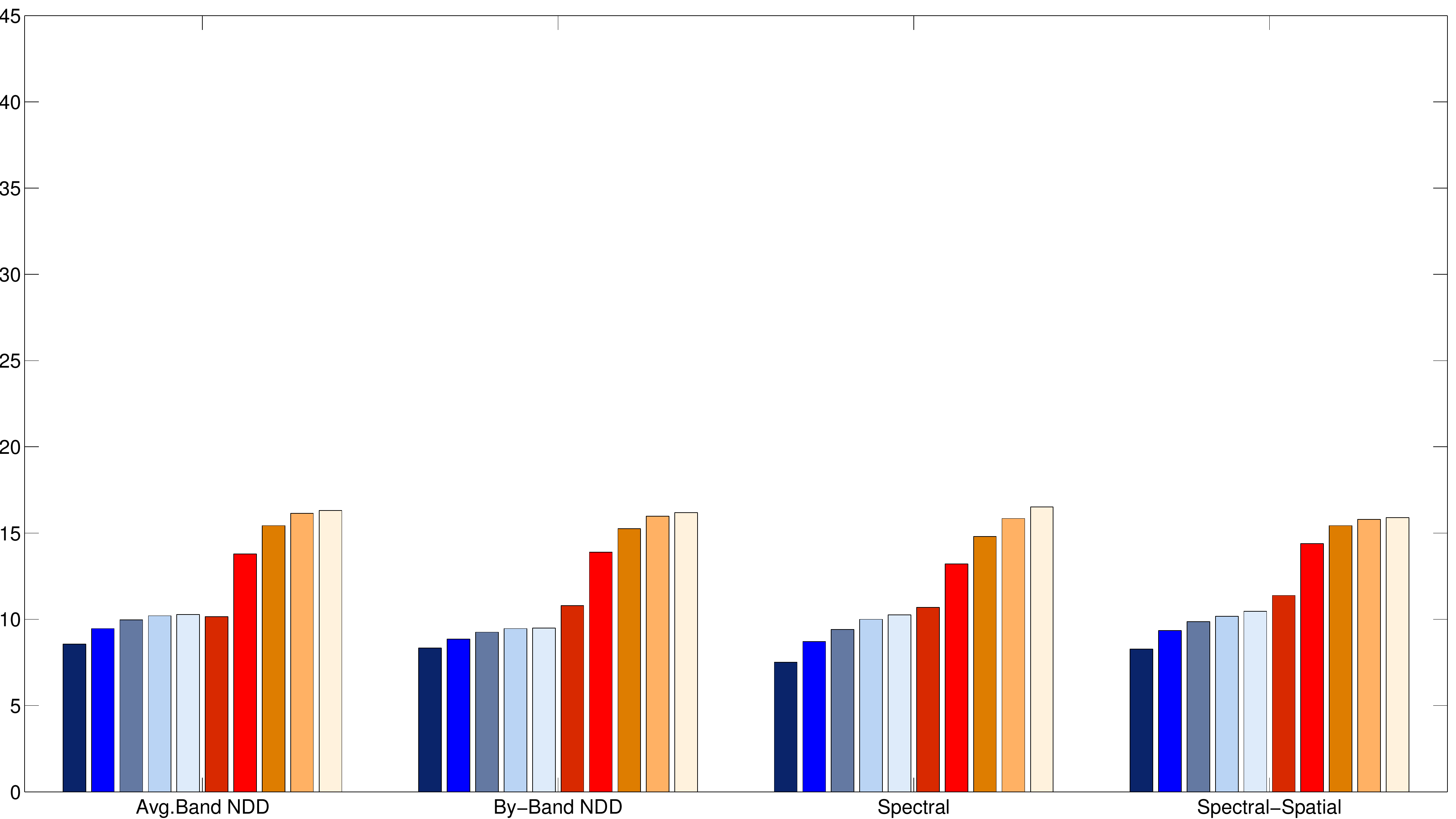} 
 & \includegraphics[width=0.4\columnwidth]{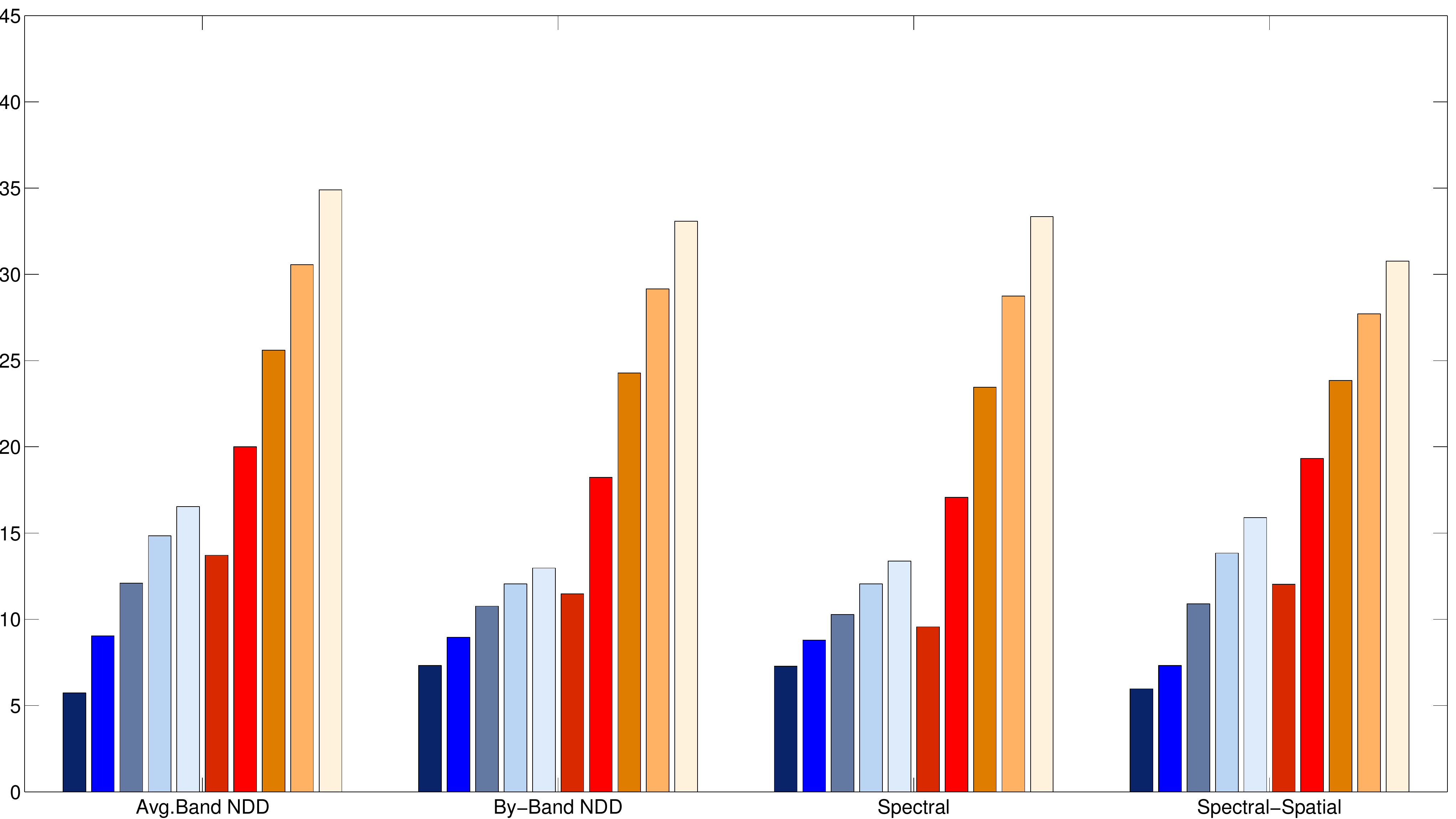} 
& \tabularnewline
\multicolumn{2}{c}{Offline Prototypes - 7NN} &
\tabularnewline
 \includegraphics[width=0.4\columnwidth]{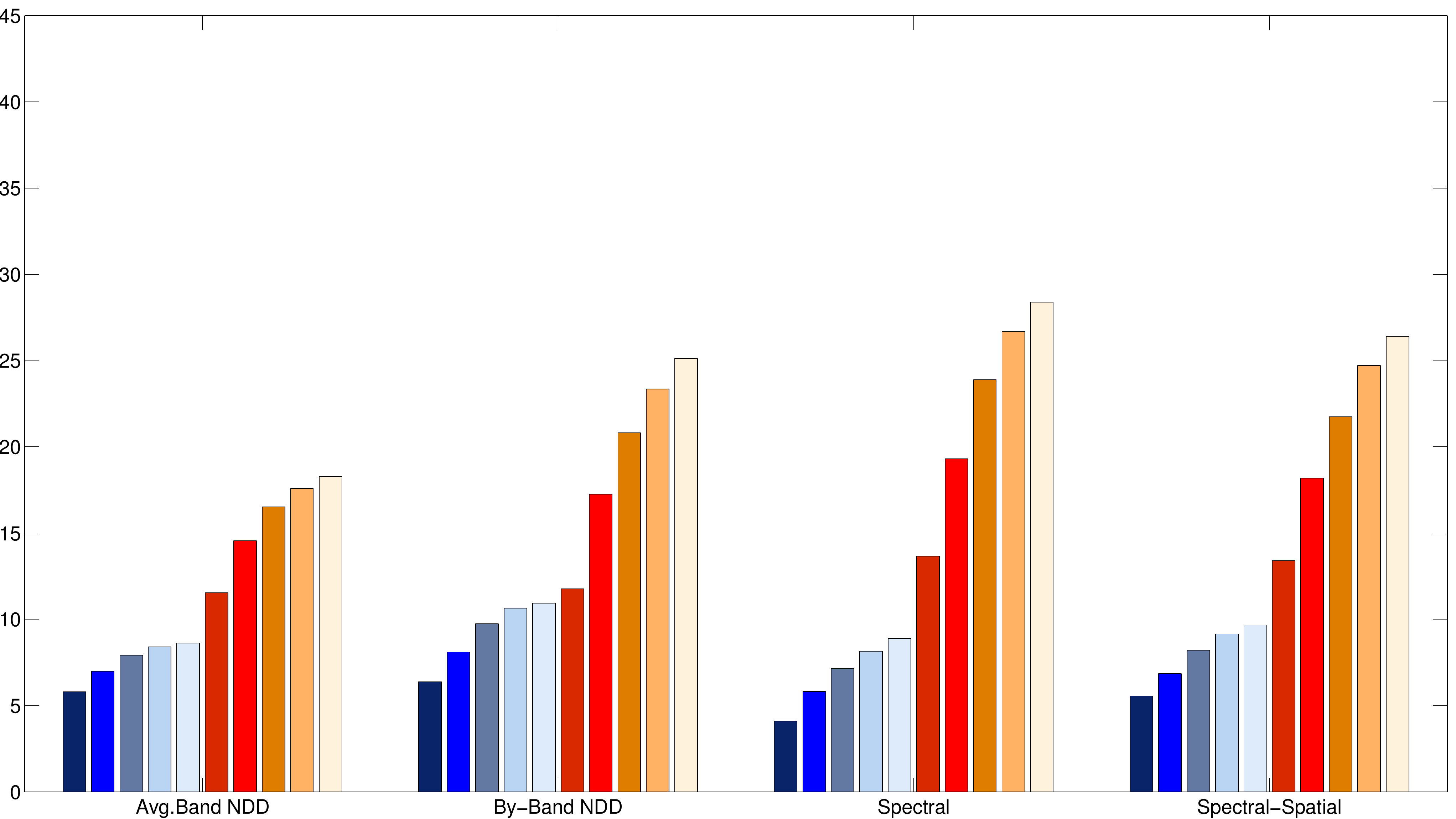} 
 & \includegraphics[width=0.4\columnwidth]{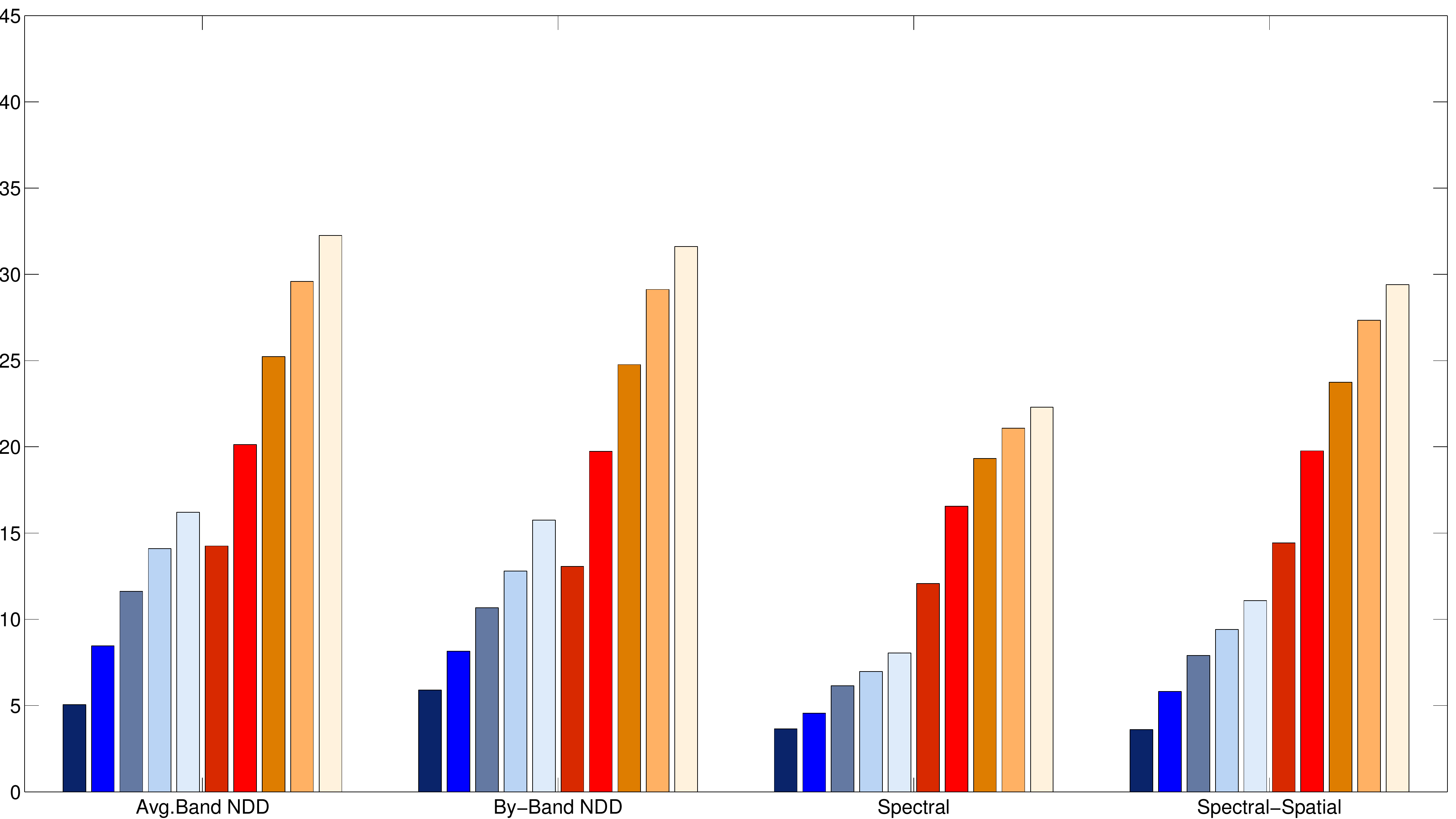} 
& \tabularnewline
\multicolumn{2}{c}{Offline Prototypes - SVM} &
\tabularnewline
\end{tabular}
\caption{\label{fig:TS_Forests}Average number of relevant (R) and non-relevant (NR) images in the training set for each RF iteration and comparing hyperspectral dissimilarity functions, using the BW and the AL image retrieval selection criteria for the Forests categorical search.}
\end{figure}

\begin{figure}[H]
\begin{centering}
\includegraphics[width=1\columnwidth]{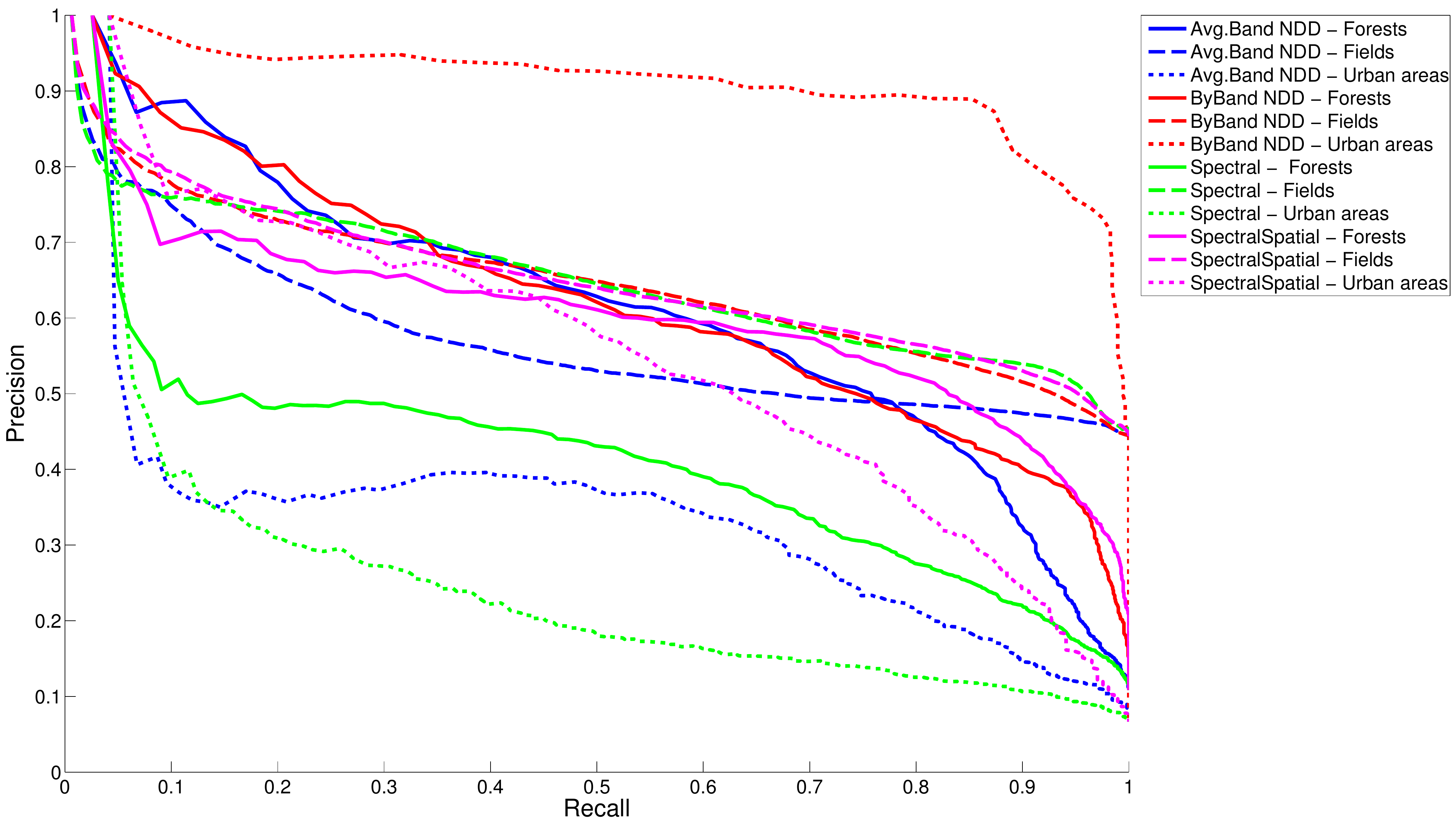} 
\par\end{centering}
\caption{\label{fig:PR_zq}Precision-Recall curves for the zero query.}
\end{figure}

\begin{figure}[H]
\begin{centering}
\includegraphics[width=1\columnwidth]{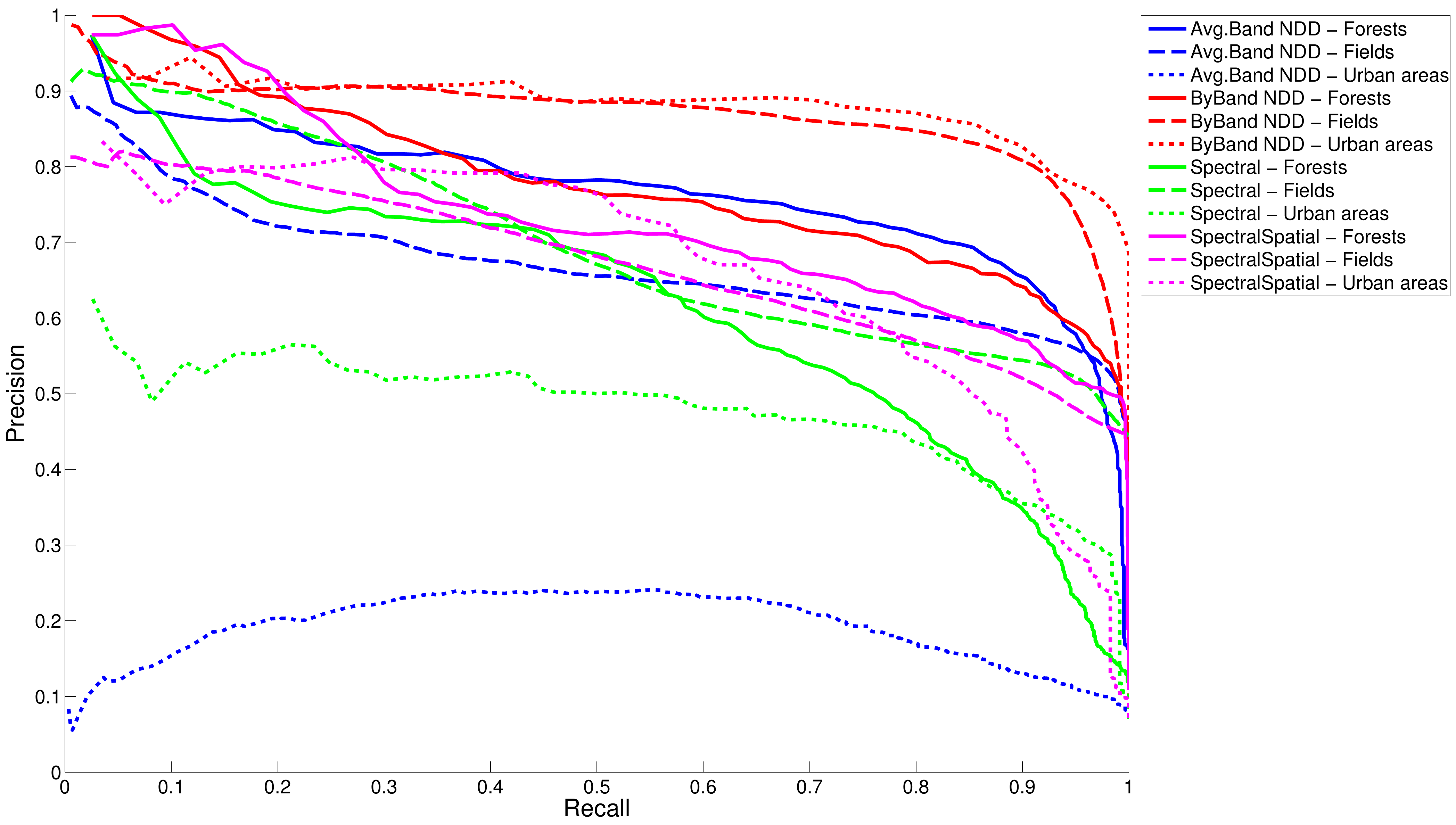} 
\par\end{centering}
\caption{\label{fig:PR_rf}Precision-Recall curves for the best RF results.}
\end{figure}

Finally, Figures \ref{fig:PR_zq} and \ref{fig:PR_rf} show the P-R curves \eqref{eq:averaged_precision} \eqref{eq:averaged recall} for the zero-query and the best RF results respectively, using the four comparing dissimilarity functions. The improve on the P-R curves by the RF process is clear except for the Urban areas categorical search, due to the pernicious effect of the lack of positive samples and the consequent asymmetrical distribution of R/NR samples on the training sets.

\section{Conclusions}
\label{sec:conclusions}

We have extended the hyperspectral CBIR systems present on the literature by a RF process based on dissimilarity spaces. To define a relevance feedback process for hyperspectral CBIR systems is not easy as most of the available hyperspectral CBIR systems rely on feature respresentations and dissimilarity functions that do not fulfil the conditions to be used in common machine learning RF processes. The proposed approach expands the available dissimilarity-based hyperspectral CBIR systems on the literature in a simple way by using dissimilarity space instead of the usual feature space. The proposed approach proved to improve the performance of the hyperspectral CBIR systems in the preliminary experiments presented on this paper. Also, the selection of a proper training set for the RF process was pointed as a major issue affecting the performance of the proposed hyperspectral RF-CBIR system. Further research will focus on this aspect and on the validation of the proposed system in a real scenario with a big database of hyperspectral images and real users.

\section*{Acknowledgements}

The authors very much acknowledge the support of Dr. Martin Bachmann
from DLR.




\bibliographystyle{plain}
\bibliography{bibliography}







\end{document}